 \documentclass[preprint,aps]{revtex4}

\usepackage{graphicx}
\usepackage{dcolumn}
\usepackage{bm}

\begin{document}


\title{Magnetization reversal in amorphous Fe/Dy multilayers: a Monte Carlo study}

\author{Etienne Talbot}
\author{Denis Ledue}%
 \email{denis.ledue@univ-rouen.fr}
\author{Pierre-Emmanuel Berche}%
\affiliation{%
Groupe de Physique des Mat\'{e}riaux, Universit\'{e} de Rouen, UMR 6634 CNRS;\\
 avenue de l'universit\'{e} - BP12 76801 Saint Etienne du Rouvray, France
}
%


\begin{abstract}
The Monte Carlo method in the canonical ensemble is used to investigate magnetization reversal in amorphous transition metal - rare earth multilayers. Our study is based on a model containing diluted clusters which exhibit an effective uniaxial anisotropy in competition with random magnetic anisotropy in the matrix. We simulate hysteresis loops for an abrupt profile and a diffuse one obtained from atom probe tomography analyses. Our results evidence that the atom probe tomography profile favors perpendicular magnetic anisotropy in agreement with magnetic measurements. Moreover, the hysteresis loops calculated at several temperatures qualitatively agree with the experimental ones. 

 \end{abstract}


\maketitle
\section{\label{sec:intro} Introduction}
In the last decades, compositionally modulated films have been extensively investigated because of their interesting magnetic properties and potential applications in magneto-optical recording. In particular, transition metal (TM) / rare earth (RE) amorphous multilayers, such as Fe/Dy multilayers, have been studied because of to their particular magnetic properties \cite{TAKA02,WANG93,SATO87,SHIN89,TAPP96,SHAN90A,SHAN90B,MIBU98,TAMI06,TAMI07}. In some conditions, they may exhibit a large magnetic anisotropy perpendicular to the layer, a high Curie temperature and a high coercive field for example. Nevertheless, the origin of the perpendicular magnetic anisotropy (PMA) is not yet clearly understood and can be explained by different models. These models are based either on an anisotropic distribution of TM-RE pairs along the perpendicular direction \cite{SATO86}, dipolar interactions \cite{SCHU95}, local structural anisotropy \cite{MERG93,FUJI96} or single ion anisotropy \cite{SHAN90B}. Recently, the results of Tamion~\textit{et al}. on (Fe 3nm/Dy 2nm) multilayers  \cite{TAMI06,TAMI07} have allowed to correlate the uniaxial anisotropy constant with the elaboration temperature by means of atom probe tomography, SQUID magnetometry and polarized neutron reflectivity measurements. It has been shown that the PMA is maximum for an elaboration temperature $T_{\mathrm{S}}=570$K, that is for diffuse interfaces. Consequently, PMA is rather due to the formation of an amorphous Fe-Dy alloy at the interfaces than abrupt interfaces.

In the present study, we investigate the influence of the concentration profile along the multilayer on the PMA of amorphous Fe/Dy multilayers by means of Monte Carlo simulations of hysteresis loops. In the framework of a local structural anisotropy model \cite{MERG93} we study the influence of the single-ion anisotropy constant on the RE atoms. Our aim is to reproduce qualitatively the experimental hysteresis loops. Thus, we are able to propose magnetization reversal mechanisms in relation with the local magnetic anisotropy and the concentration profile. Our numerical results also provide complementary information such as magnetization profile along the multilayers for different values of the applied field. The model and simulation technique are described in Sec.~\ref{sec:model}. In the following sections, we report and discuss numerical results on the influence of the concentration profile at low temperature for different types of magnetic anisotropy: in the case of uniaxial anisotropy on all Dy sites in Sec.~\ref{sec:uniax}, for the cluster anisotropy model (described below) in Sec.~\ref{sec:amas} and in the case of the cluster anisotropy model combined with random anisotropy in Sec.~\ref{sec:amasrma}. More precisely, Sec.~\ref{sec:uniax} is devoted to a simple model for which theoretical results are available which has allowed us to validate the simulation technique. The investigation, in Sec.~\ref{sec:amas}, of the cluster anisotropy model is performed to understand the differences with the uniaxial anisotropy model. The results of this section are very useful to explain the numerical data obtained in Sec.~\ref{sec:amasrma}. In Sec.~\ref{sec:disc}, we report the temperature effect on hysteresis loops and compare our results to experimental ones. Finally a conclusion is given in Sec.~\ref{sec:concl}.

\section{\label{sec:model} Model and Monte Carlo simulation}

\subsection{Description of the model}

Our model consists in a face centered cubic (FCC) multilayer system made up of Fe and Dy atoms. We choose this closed-packed structure (each atom has 12 nearest neighbors) because its density is very close to those of amorphous structures \cite{HANS89,HEIM76,HOND94}. To take into account the different interatomic distances observed in an amorphous material, we consider distributed exchange interactions since they strongly depend on the interatomic distances \cite{HAND69}.

\begin{figure}\centering
\rotatebox{-90}{\includegraphics[width=0.3\textwidth]{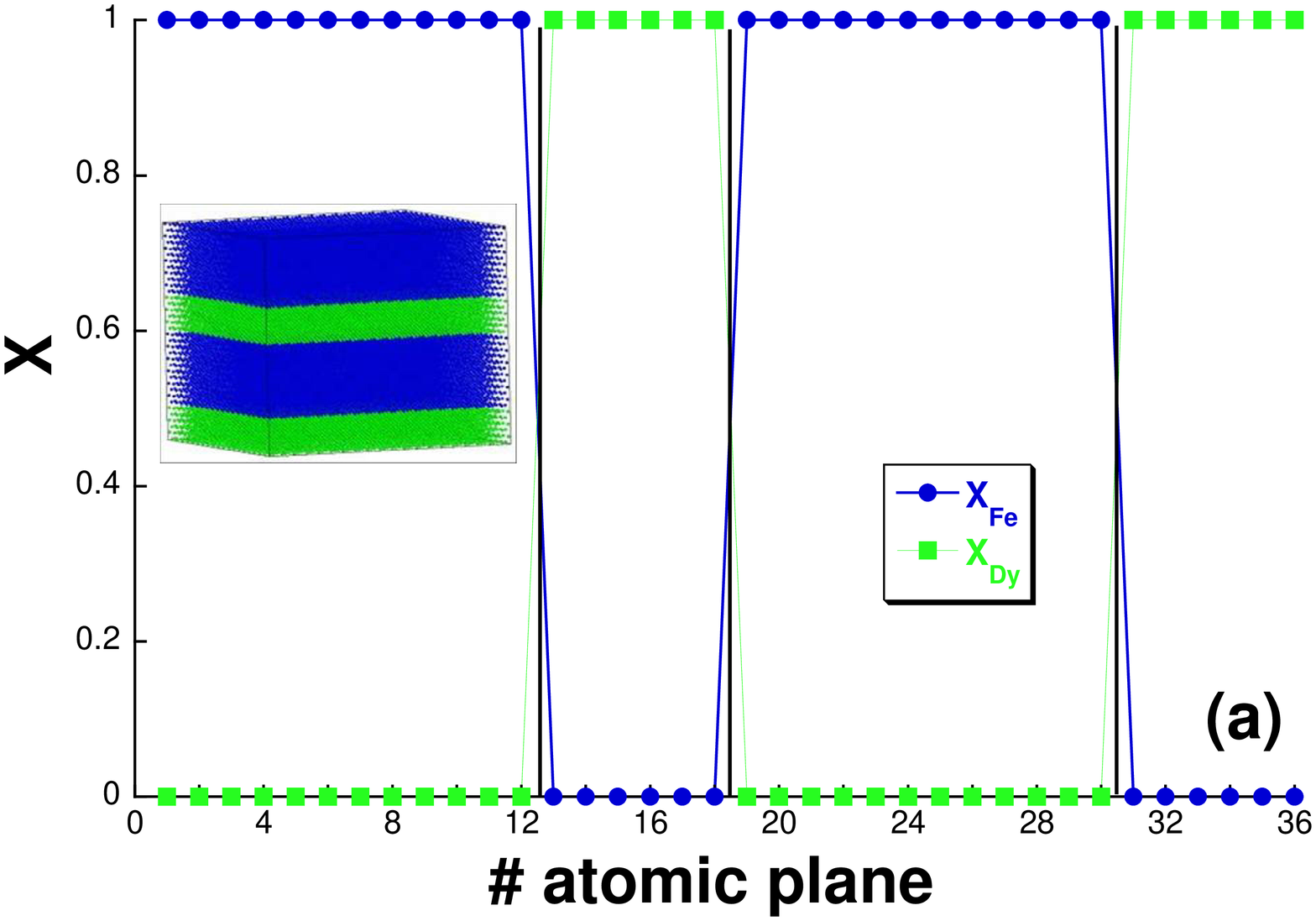}}
\rotatebox{-90}{\includegraphics[width=0.3\textwidth]{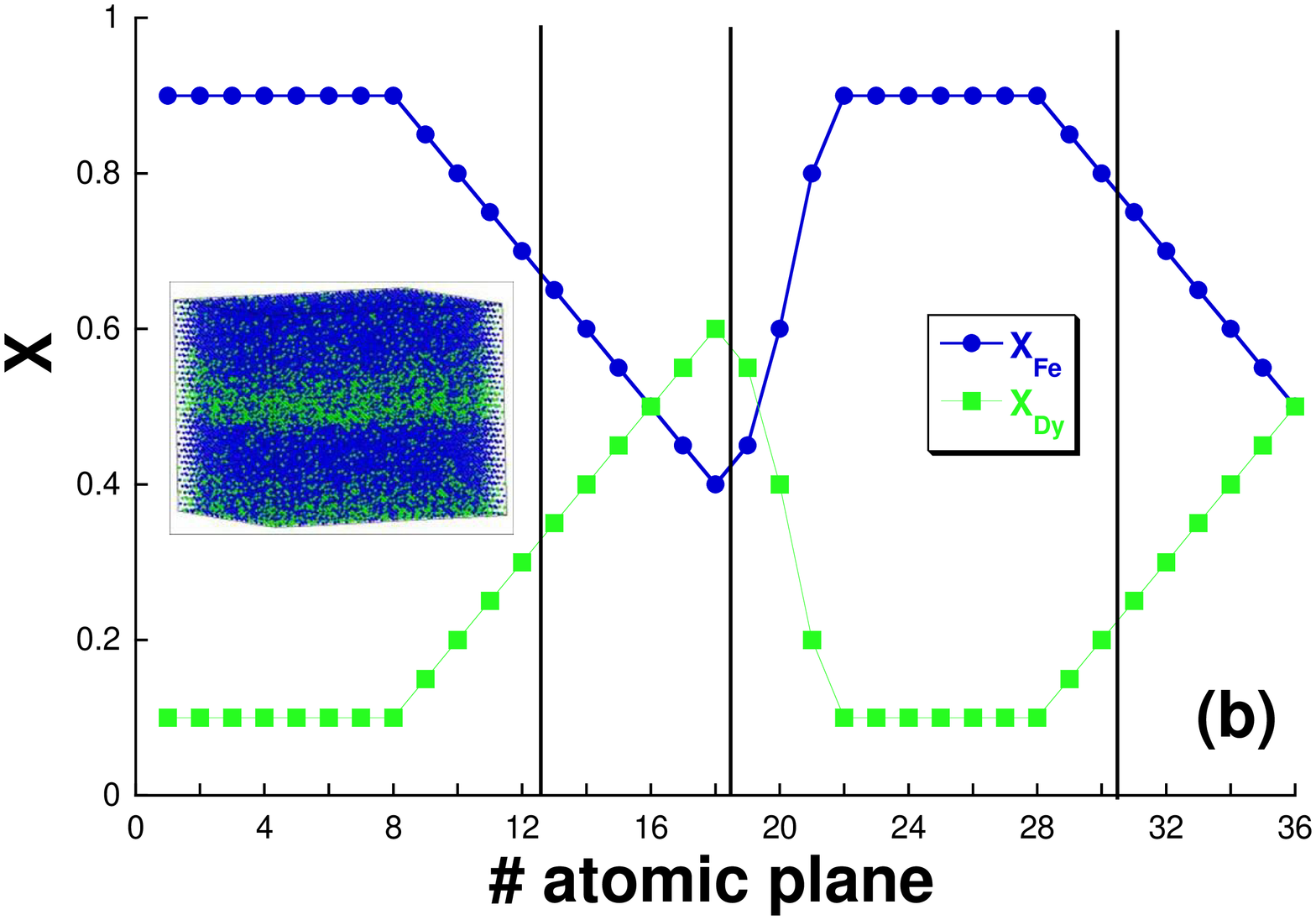}}
\caption{Concentration profile of each specie along the multilayer growth direction for an \textit{abrupt} (a) and a \textit{APT} (b) concentration profile.}
\label{figure1}

\end{figure}

It has been shown previously that atomic diffusion has a majour influence on the macroscopic magnetic anisotropy \cite{TAMI06}. Consequently, two different concentration profiles are used to explain these features. The first one, called abrupt in the following, corresponds to a multilayer made up of pure Fe and Dy layers with abrupt interfaces (Fig.~\ref{figure1}.(a)). The second profile has been directly obtained from atom probe tomography analysis of (Fe 3nm/Dy 2nm) multilayers elaborated at 570K \cite{TAMI06}; it is called APT (atom probe tomography) profile in the following. This profile is composed of a Fe-rich region (Fe$_{90}$Dy$_{10}$) and a region in which the concentration varies; we would like to note that this profile does not display any pure region (Fig.~\ref{figure1}.(b)). As it has been previously mentioned, this concentration profile leads experimentally to the maximum of PMA. In our model, there are 36 atomic planes corresponding to a double (Fe 3nm/Dy 2nm) bilayer. For simplicity, we do not take into account the difference in atomic radius of Fe and Dy atoms. We apply periodic boundary conditions in the film plane and free boundary conditions in the $z$-direction. In sections~\ref{sec:amasrma} and \ref{sec:disc}, in order to compare with experimental results, we calculate the physical quantities over the central bilayer only, to get rid off free surface effects.

Each node of the FCC lattice is occupied by a classical Heisenberg spin which is a 3D vector spin. This classical model is the most suitable for investigating magnetic configurations in systems with competitive interactions (exchange, anisotropy and Zeeman). We consider the following Hamiltonian:

\begin{equation}
	E = - \sum_{<i,j>} J_{ij} (\mathbf{S}_{i}.\mathbf{S}_{j}) + E_{\mathrm{a}} - \mathbf{B} \sum_{i} \mathbf{m}_{i},
	\label{eq:hamilt}
\end{equation}
where $\mathbf{S}_{i}$ is a classical Heisenberg spin and $\mathbf{m}_{i}$ is the magnetic moment at site $i$. The first term corresponds to exchange energy. Here, we consider nearest-neighbor (NN) interactions only, so $J_{ij}$ are the NN exchange interactions. The second term, $E_{\mathrm{a}}$, is the magnetic anisotropy energy. We consider only single-ion anisotropy on the Dy atoms, the magnetic anisotropy of the Fe atoms is neglected. The third term  describes the Zeeman energy where $\mathbf{B}$ is the applied magnetic field. For the Fe spins, we take the values proposed by Heiman~\textit{et al}. \cite{HEIM76} which depend on the local concentration under the form:

\begin{equation}
	S_{\mathrm{Fe}} (X_{\mathrm{Fe}}) = 1.1-1.125(1-X_{\mathrm{Fe}}) \quad (X_{\mathrm{Fe}}>0.4).
	\label{eq:spinFe}
\end{equation}
Since the Dy spin is not sensitive to the local environment, we take $S_{\mathrm{Dy}}=2.5$. The Land\'{e} factors are $g_{\mathrm{Fe}}=2$ and $g_{\mathrm{Dy}}=4/3$. The atomic moments for Fe and Dy are related to the spins according to:

\begin{equation}
m_{\mathrm{Fe}} = g_{\mathrm{Fe}} \mu_{\mathrm{B}} S_{\mathrm{Fe}} (X_{\mathrm{Fe}}),
	\label{eq:m_Fe}
\end{equation}
\begin{equation}
	m_{\mathrm{Dy}} = \frac{g_{\mathrm{Dy}}}{g_{\mathrm{Dy}}-1} \mu_{\mathrm{B}} S_{\mathrm{Dy}} = 10 \mu_{\mathrm{B}}.
	\label{eq:m_Dy}
\end{equation}

 The NN exchange interactions are also extracted from the work of Heiman~\textit{et al.} \cite{HEIM76} on Fe-Dy amorphous thin films. Since they have been obtained by mean-field calculations, we have to adjust them to get the pure amorphous Fe and Dy Curie temperatures (respectively 270K \cite{HEIM76,TAPP96} and 110K \cite{TAPP96}) by Monte Carlo simulations. 

The Fe-Fe and Fe-Dy exchange interactions linearly depend on the local concentration $X_{\mathrm{Fe}}$. The Dy-Dy exchange interaction, which is much smaller than the others, is independent of the concentration. These exchange interactions are given by:
\begin{equation}
	J_{\mathrm{Fe-Fe}}(X_{\mathrm{Fe}})/k_{\mathrm{B}} = 77 + 449(1-X_{\mathrm{Fe}}) \quad (\text{in K}),
	\label{eq:Jfefe}
\end{equation}
\begin{equation}
	J_{\mathrm{Fe-Dy}}(X_{\mathrm{Fe}})/k_{\mathrm{B}} = 8 - 198(1-X_{\mathrm{Fe}}) \quad (\text{in K}),
	\label{eq:Jfedy}
\end{equation}
\begin{equation}
	J_{\mathrm{Dy-Dy}}/k_{\mathrm{B}} = 6.5 \quad (\text{in K}).
	\label{eq:Jdydy}
\end{equation}
$J_{\mathrm{Fe-Dy}}$ is strongly negative; it is thus responsible for the ferrimagnetic order (with Fe moments antiparallel to Dy moments) which has been experimentally observed by polarized neutron reflectivity measurements \cite{TAMI07}. 

In the following, we define the ferrimagnetic magnetization per atom ($M_{\mathrm{ferrimagnetic}}$) by:
\begin{equation}
	M_{\mathrm{ferrimagnetic}} = \frac{1}{N_{p}} \sum_{l=1}^{N_{p}} (1-X_{\mathrm{Fe}}^{l})m_{\mathrm{Dy}} - X_{\mathrm{Fe}}^{l}m_{\mathrm{Fe}}^{l},	\label{eq:m_ferri}
\end{equation}
where $N_{p}$ is the number of atomic planes and $m_{\mathrm{Fe}}^{l}$ is the Fe atomic moment in the $l^{\mathrm{th}}$ plane. For the abrupt and APT profile, $M_{\mathrm{ferrimagnetic}}$ is equal to $1.87 \mu_{\mathrm{B}}$ and $1.21 \mu_{\mathrm{B}}$ respectively.


\subsection{Monte Carlo simulation}

In this work, we use Monte Carlo (MC) method in the canonical ensemble based on the Metropolis algorithm \cite{METR53}. This famous method is known to ensure a good convergence towards the thermodynamic equilibrium but it may in some circumstances not describe properly the magnetization reversal. So we use here the algorithm proposed by Nowak~\textit{et al.} \cite{NOWA00}. In this algorithm, a site $i$ is chosen randomly and the new orientation of the spin of this site within a cone with a given size is proposed. To achieve this procedure, a random vector $\mathbf{u}$ with uniform probability distribution within a sphere of radius R$<1$ is added to the initial normalized spin $\mathbf{s}_{i} = \mathbf{S}_{i}/ \parallel \mathbf{S}_{i} \parallel$. The new orientation of the spin is then given by the unit vector:

\begin{equation}
\mathbf {s}^{'}_{i}=\frac{\mathbf {s}_{i}+\mathbf {u}}{\parallel\mathbf {s}_{i}+\mathbf {u}\parallel}.
\end{equation}

The energy difference $\Delta E$ between the new ($\mathbf {S}^{'}_{i}$) and initial ($\mathbf {S}_{i}$) spin orientations is calculated from Eq.(\ref{eq:hamilt}) (with $\mathbf{S}^{'}_{i}= \parallel \mathbf{S}_{i} \parallel \mathbf{s}^{'}_{i}$ ). If $\Delta E<0$, the transition is accepted since it lowers the energy. Otherwise, the transition is accepted with a probability rate proportional to $\text{exp}(-\Delta E / k_{\mathrm{B}}T)$. One Monte Carlo step (MCS) consists in examining all spins of the system once. The variation of R allows varying the acceptance rate in order to optimize the efficiency of the algorithm. With this technique, reliable MC simulations with a reasonable number of MCS can be performed at low temperature thanks to large enough acceptance rates. Moreover, unlike the standard Metropolis algorithm, this method prevents non physical spin flips by tunneling accross the barrier.

The initial magnetic configuration is completely disordered. Then, a strong external field $+\mathbf{B}_{\text{max}}$ is applied along a given axis. The hysteresis loop is performed by varying the external field with a step $\Delta B=0.15$T. At each field, 40 000 MCS are discarded for local equilibration before averaging the magnetization components over the next 40 000 MCS. As we are interested in multilayers with configurational disorder, the final hysteresis loops are measured by averaging over several disorder configurations.

In order to choose reasonable values of R, we perform numerical simulations on simple systems which satisfy the Stoner-Wohlfarth (SW) model \cite{STON48}. Then, we keep for R the values which allow to reproduce the theoretical hysteresis loops within a reasonable computational time (R $\sim 0.1$).

\section{\label{sec:uniax}Uniaxial anisotropy}
In this section, we consider uniaxial anisotropy perpendicular to the film plane for all Dy atoms. The anisotropy energy is given by:

\begin{equation}
  E_{\mathrm{a}}=-D_{\mathrm{Dy}} \sum_{i \in \text{Dy sublattice}} (\mathbf{S}_{i}.\mathbf{z})^{2},
	\label{eq:Ea_uniax}
\end{equation}  
where $D_{\mathrm{Dy}}$ is the single-ion anisotropy constant. We investigate here the influence of the concentration profile and of the anisotropy constant on the magnetization reversal of an amorphous Fe/Dy multilayer.

\subsection{Abrupt profile}

\begin{figure}\centering
\rotatebox{-90}{\includegraphics[width=0.3\textwidth]{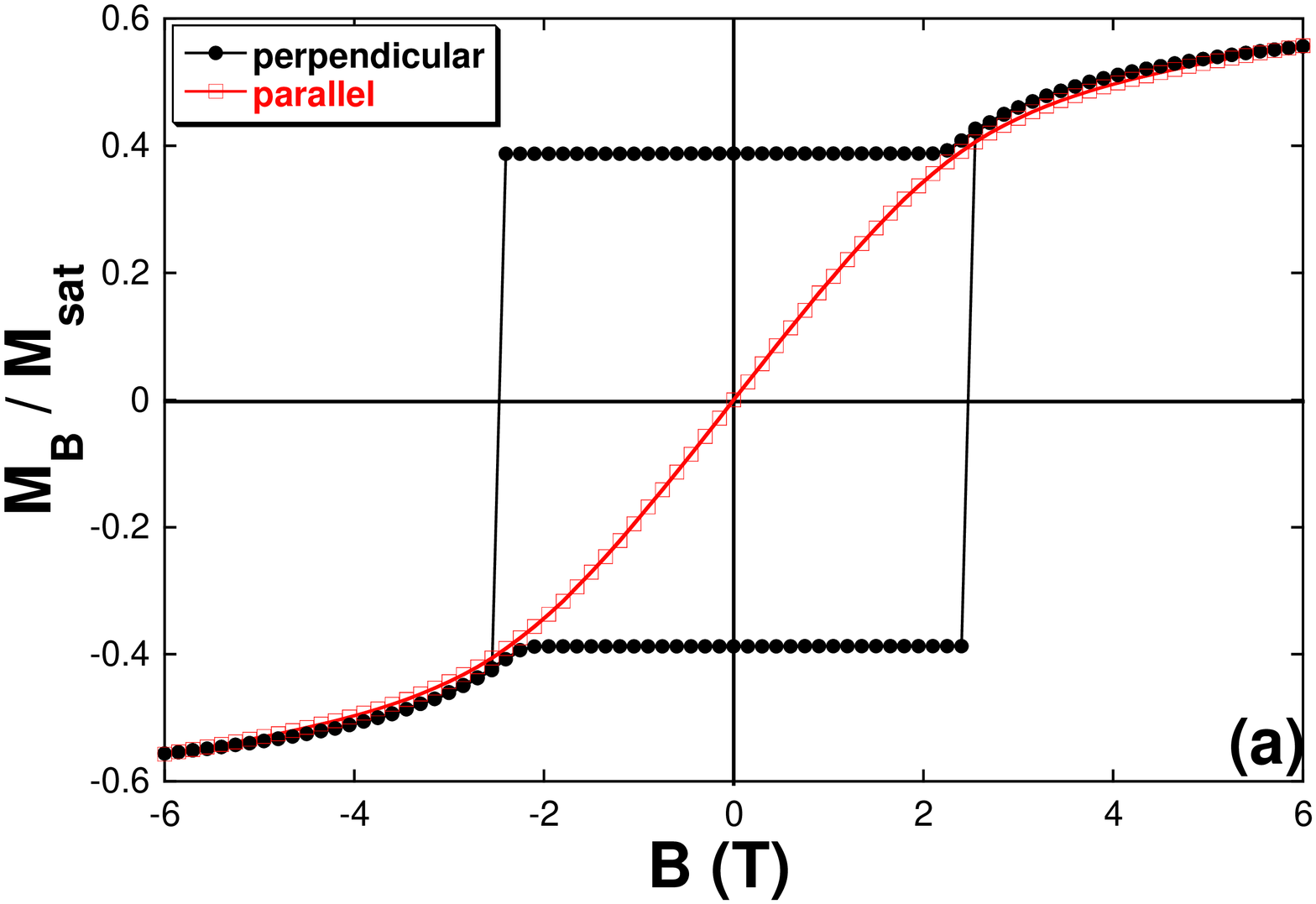}}
\rotatebox{-90}{\includegraphics[width=0.3\textwidth]{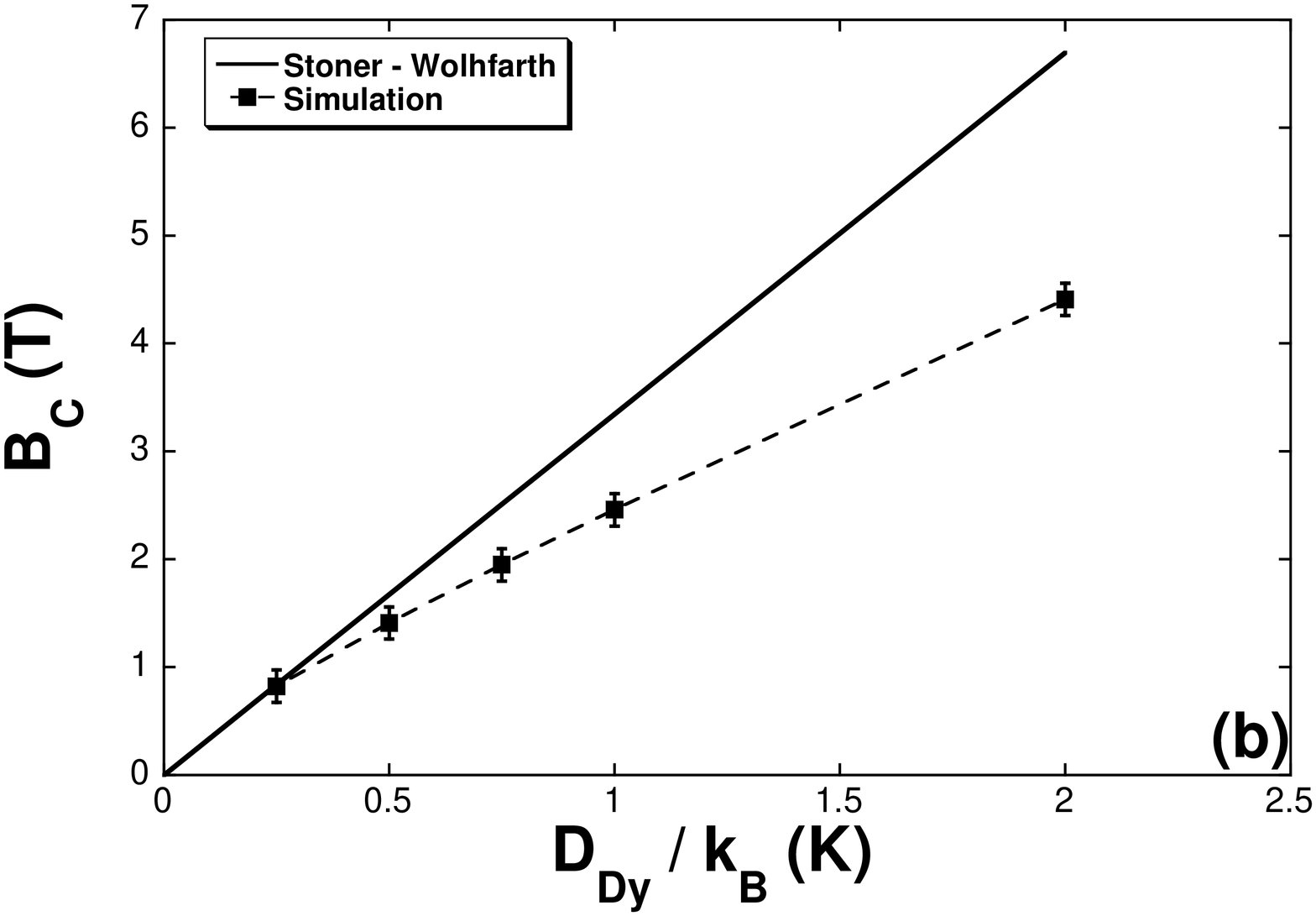}}

\caption{(a) Hysteresis loops in perpendicular and parallel orientations for an Fe/Dy multilayer with the abrupt profile and uniaxial magnetic anisotropy at $T=1$K for $D_{\mathrm{Dy}} / k_{\mathrm{B}} = 1$K. (b) Coercive field as a function of the Dy uniaxial anisotropy constant at $T=1$K.}
\label{hyst_uniax_ab}
\end{figure}

The hysteresis loops of the multilayer with an abrupt profile at low temperature for an applied field in the film plane and perpendicular to it are shown in Fig.~\ref{hyst_uniax_ab}.(a) . For the latter case, the hysteresis loop reveals two types of behaviors. For $\vert B\vert < B_{\mathrm{C}}=2,45 \pm 0,15$T, the magnetic configuration is ferrimagnetic (Fe and Dy layers are antiparallel to each other). Beyond $B_{\mathrm{C}}$, the absolute value of the magnetization increases with the external field. These configurations exhibit an interface domain wall (IDW) located in the Fe layer which has a free surface. This is due to the competition between the external field and the antiparallel Fe-Dy coupling. It has to be noted that an applied field of 6T is not strong enough to saturate the magnetization of the sample. For an in-plane field, the loop is closed and the absolute value of the magnetization increases monotonously with the field intensity as expected for a hysteresis loop measured in a hard orientation.

The switching field predicted by the SW model, i.e. for uniform rotation process, is:
\begin{equation}
	B_{\mathrm{C}}^{\mathrm{SW}}=\frac{2X_{\mathrm{Dy}}D_{\mathrm{Dy}}S_{\mathrm{Dy}}^{2}}{M_{\mathrm{ferrimagnetic}}},
	\label{eq:bcuni}
\end{equation}
where $X_{\mathrm{Dy}}$ is the Dy atomic fraction. In Fig.~\ref{hyst_uniax_ab}.(b), we have plotted the coercive field as a function of the Dy anisotropy constant for a perpendicular applied field. The simulated coercivity increases as expected with $D_{\mathrm{Dy}}$, but it is smaller than the SW values except for $D_{\mathrm{Dy}}/k_{\mathrm{B}} \leq 0.25$K. This deviation from the SW model confirms that the magnetization reversal is not uniform. The reason is that the antiferromagnetic Fe-Dy interactions favor the formation of the IDW during the reversal. Moreover, we note the existence of a spin-flop transition at $B_{\mathrm{SF}}\sim2,1$T for $D_{\mathrm{Dy}}/k_{\mathrm{B}}>1$K in good agreement with previous studies on ferrimagnetic multilayers \cite{DANT04,WORL04}. The spin-flop transition corresponds to the formation of a twisted spin structure at the interface.

\subsection{APT profile}
\begin{figure}\centering
\rotatebox{-90}{\includegraphics[width=0.3\textwidth]{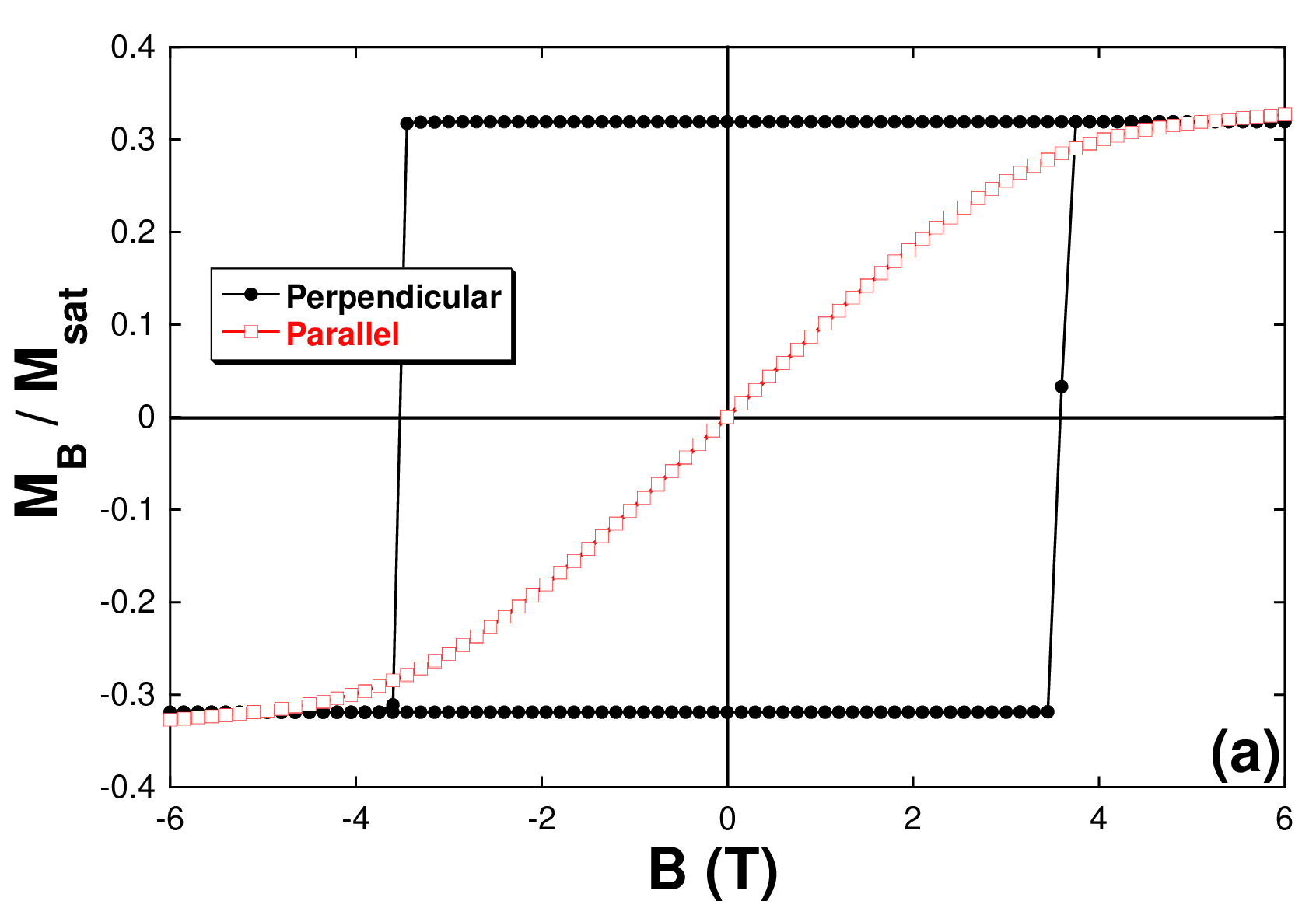}}
\rotatebox{-90}{\includegraphics[width=0.3\textwidth]{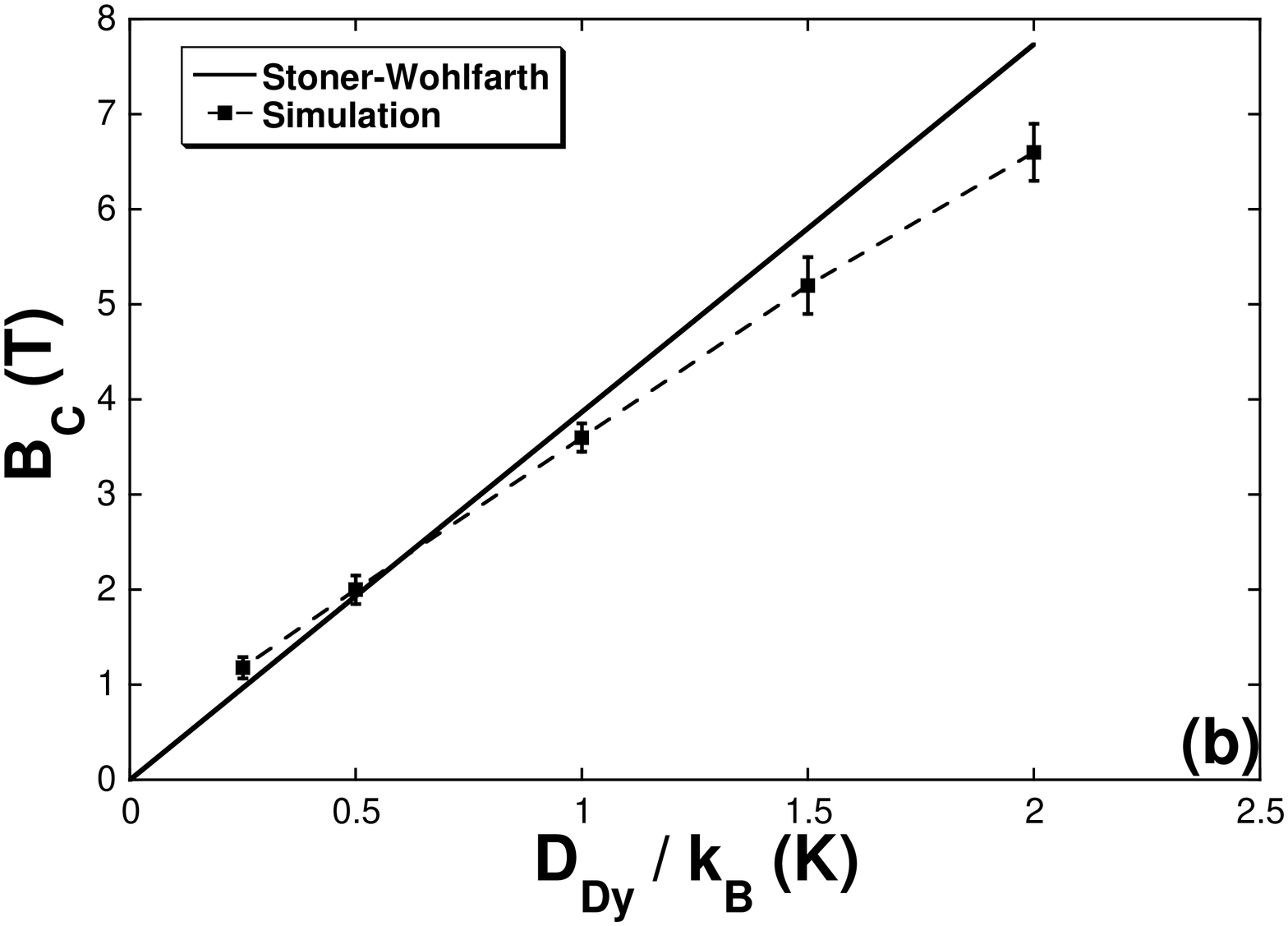}}
\caption{(a) Hysteresis loops in perpendicular and parallel orientations for an Fe/Dy multilayer with the APT profile and uniaxial magnetic anisotropy at $T=1$K for $D_{\mathrm{Dy}} / k_{\mathrm{B}} = 1$K. (b) Coercive field as a function of the Dy uniaxial anisotropy constant at $T=1$K.}
\label{hyst_uniax_so}
\end{figure}

In the case of the APT profile, the hysteresis loops at $T=1$K are shown in Fig.~\ref{hyst_uniax_so}.(a). Contrary to the previous case, we observe a square loop when the field is normal to the layers which characterizes a ferrimagnetic order for each value of the magnetic field. These results show that for the APT profile the system is equivalent to a homogeneous ferrimagnetic material (at least in the field range investigated).
In Fig.~\ref{hyst_uniax_so}.(b), we have plotted the coercivity at $T=1$K as a function of the anisotropy constant $D_{\mathrm{Dy}}$. We obtain a reasonable agreement with the SW model up to $D_{\mathrm{Dy}} / k_{\mathrm{B}} \sim 1$K, i.e. the magnetization reversal is roughly uniform. For $D_{\mathrm{Dy}} / k_{\mathrm{B}} >1$K, our results deviate significantly from the SW model. Indeed, the magnetization reversal starts in Fe-rich planes by breaking Fe-Dy antiparallel couplings which leads to a non-uniform rotation process. Then, we can conclude that the APT profile favors uniform rotation process in comparison with the abrupt profile.

\subsection{Influence of the temperature}
\begin{figure}\centering
\rotatebox{-90}{\includegraphics[width=0.3\textwidth]{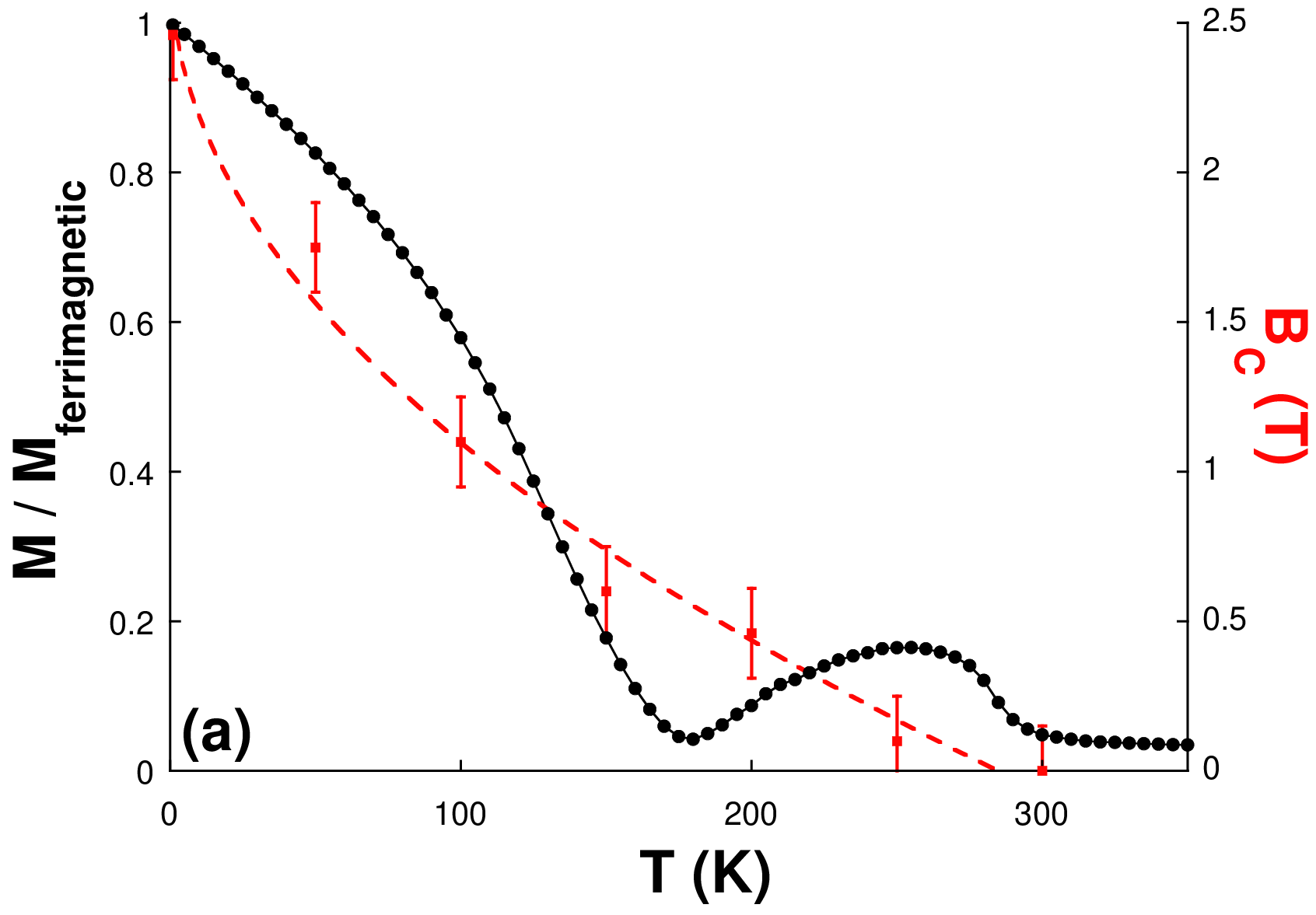}}
\rotatebox{-90}{\includegraphics[width=0.3\textwidth]{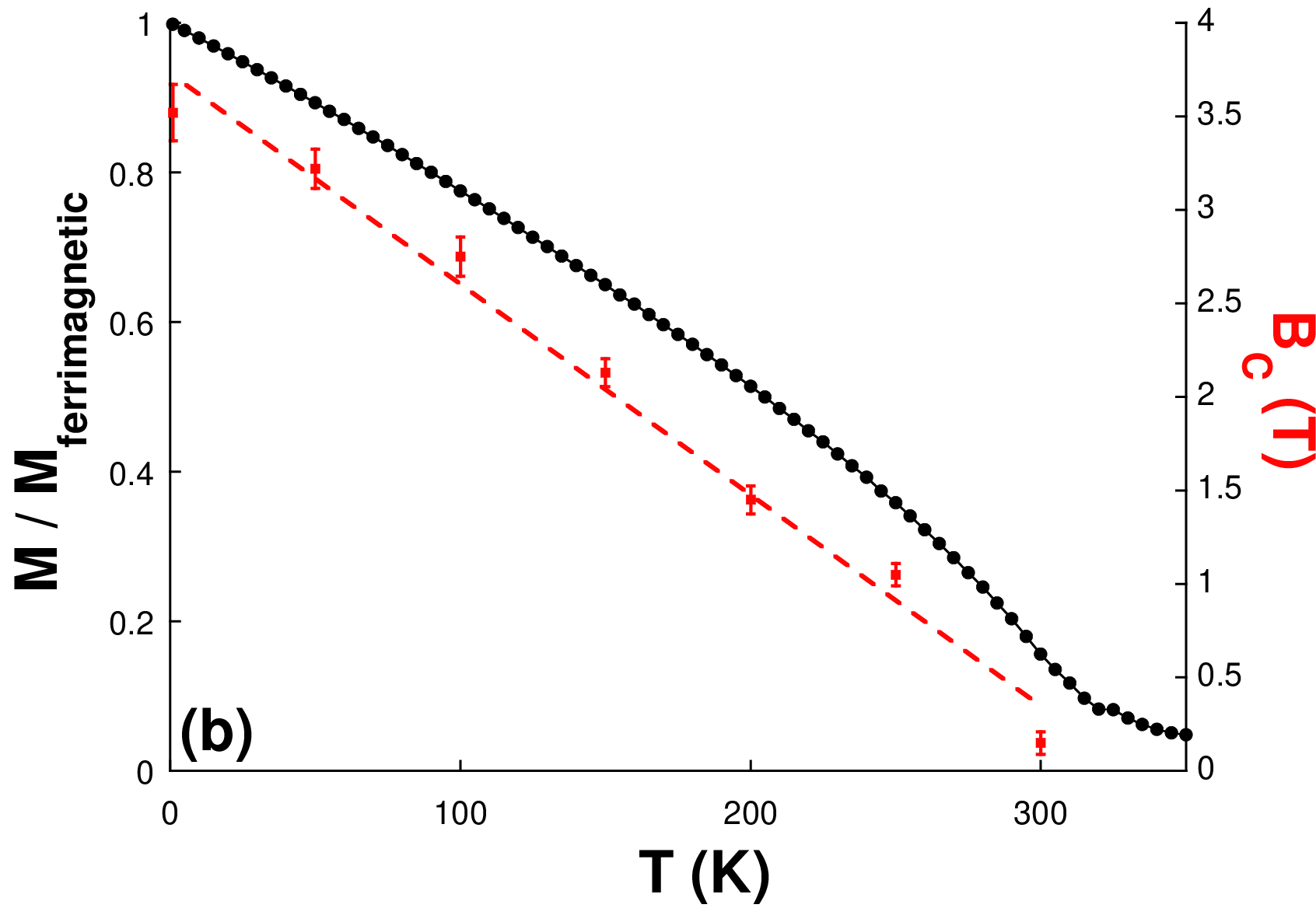}}

\caption{Thermal variation of the reduced magnetization (continuous line) and of the coercive field (dashed line) for $D_{\mathrm{Dy}} / k_{\mathrm{B}}=1$K ((a) abrupt profile; (b) APT profile).}
\label{coerciv_temp}
\end{figure}

Finally, we have studied the temperature influence on the hysteresis loops for each concentration profile. The thermal variation of the reduced magnetization and of the coercive field is shown in Fig.~\ref{coerciv_temp}.

Our results evidence a faster decrease of the coercivity as the temperature increases in the case of the abrupt profile (Fig.~\ref{coerciv_temp}). The temperature dependence of the coercivity can be fitted to a square law contrary to the APT profile for which the decrease is roughly linear. This faster decrease is related to the decrease of the magnetization which is also faster than for the APT profile \cite{TALB08}.

\section{\label{sec:amas} Cluster anisotropy }
\begin{figure}\centering
{\includegraphics[width=0.35\textwidth]{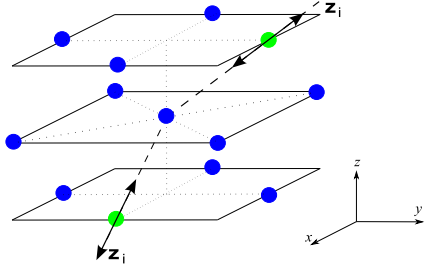}}
\caption{Schematic representation of the ordered clusters. Fe atoms are in blue,
Dy atoms in green. The arrows show the local anisotropy axes on the Dy sites.}
\label{schema_amas}
\end{figure}

Here, we consider the local anisotropy model. This model is based on the existence of small crystallized clusters at the scale of a few interatomic distances depending on the elaboration method. These clusters define on average an easy axis perpendicular to the film plane \cite{MERG93}. In our model, they consist of 13 atoms (one Fe central atom and its 12 nearest neighbors) (Fig.~\ref{schema_amas}). Among the 8 neighbors which are not in the $xy$ plane of the central atom, between 2 and 4 Dy atoms are randomly distributed in order to obtain a cluster composition close to that of defined compounds. The easy axes (on Dy atoms belonging to the clusters) are along the Fe-Dy bonds; so the unit vectors $\mathbf{z}_{i}$ of these 4 axes are $(\pm \frac{1}{\sqrt{2}}$,0,$\frac{1}{\sqrt{2}}$) or ($0,\pm \frac{1}{\sqrt{2}}$,$\frac{1}{\sqrt{2}}$). This leads on average to an easy axis perpendicular to the film plane.

 In this case, the anisotropy energy is defined as:
\begin{equation}
	E_{\mathrm{a}}=-D_{\mathrm{Dy}} \sum_{i \in \text{Dy in clusters}} (\mathbf{S}_{i}.\mathbf{z}_{i})^{2}.
	\label{eq:Ea_clust}
\end{equation}  
As the anisotropy coefficient $D_{\mathrm{Dy}}$ in amorphous multilayers is not accurately determined in the literature, we consider it as a free parameter. In the same way, the cluster concentration ($c_\mathrm{cluster}$), which is defined as the number of atoms (Dy and Fe) included in the clusters divided by the total number of atoms, is a free parameter in the simulations since it cannot be evaluated experimentally.

We investigate the influence of the Dy anisotropy constant and of the cluster concentration for the two concentration profiles. The cluster concentration varies from 5\% to 20\%, that is only 3.4\% to 13.8\% of the Dy atoms display single-ion anisotropy. Consequently, we take larger values of the anisotropy constant than in the previous section in order to obtain significant values of the coercive field. In the framework of the SW model, it can easily be shown that the coercive field is given by: 
\begin{equation}
B_{\mathrm{C}}^{\mathrm{SW}}=\frac{X_{\mathrm{Dy}}^{\mathrm{cluster}}D_{\mathrm{Dy}}S_{\mathrm{Dy}}^{2}}{2M_{\mathrm{ferrimagnetic}}},	
	\label{eq:bc_clus}
\end{equation}
where $X_{\mathrm{Dy}}^{\mathrm{cluster}}$ is the atomic fraction of Dy atoms belonging to the clusters.

\subsection{Abrupt profile}
\begin{figure}\centering
\rotatebox{-90}{\includegraphics[width=0.3\textwidth]{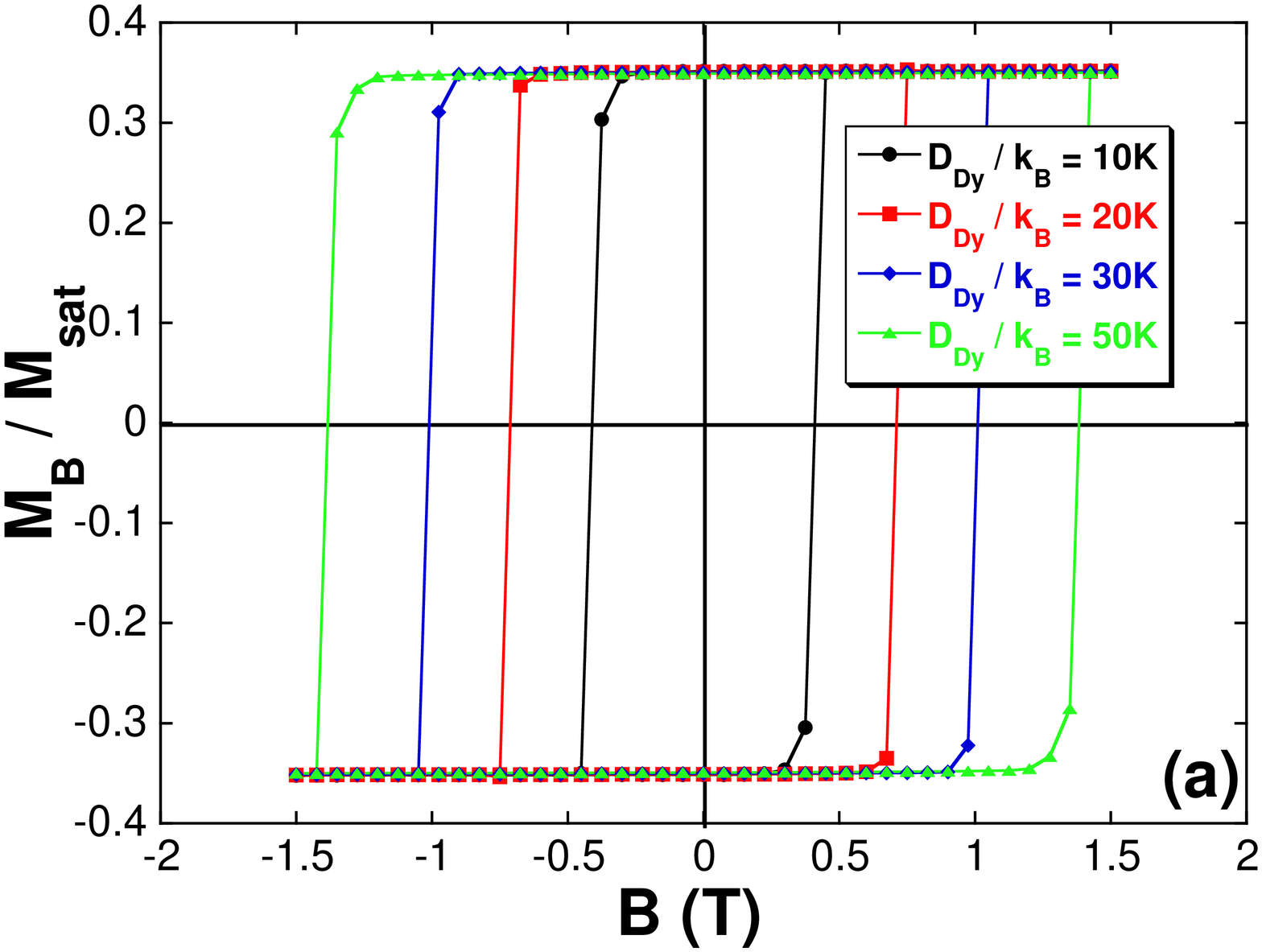}}
\rotatebox{-90}{\includegraphics[width=0.3\textwidth]{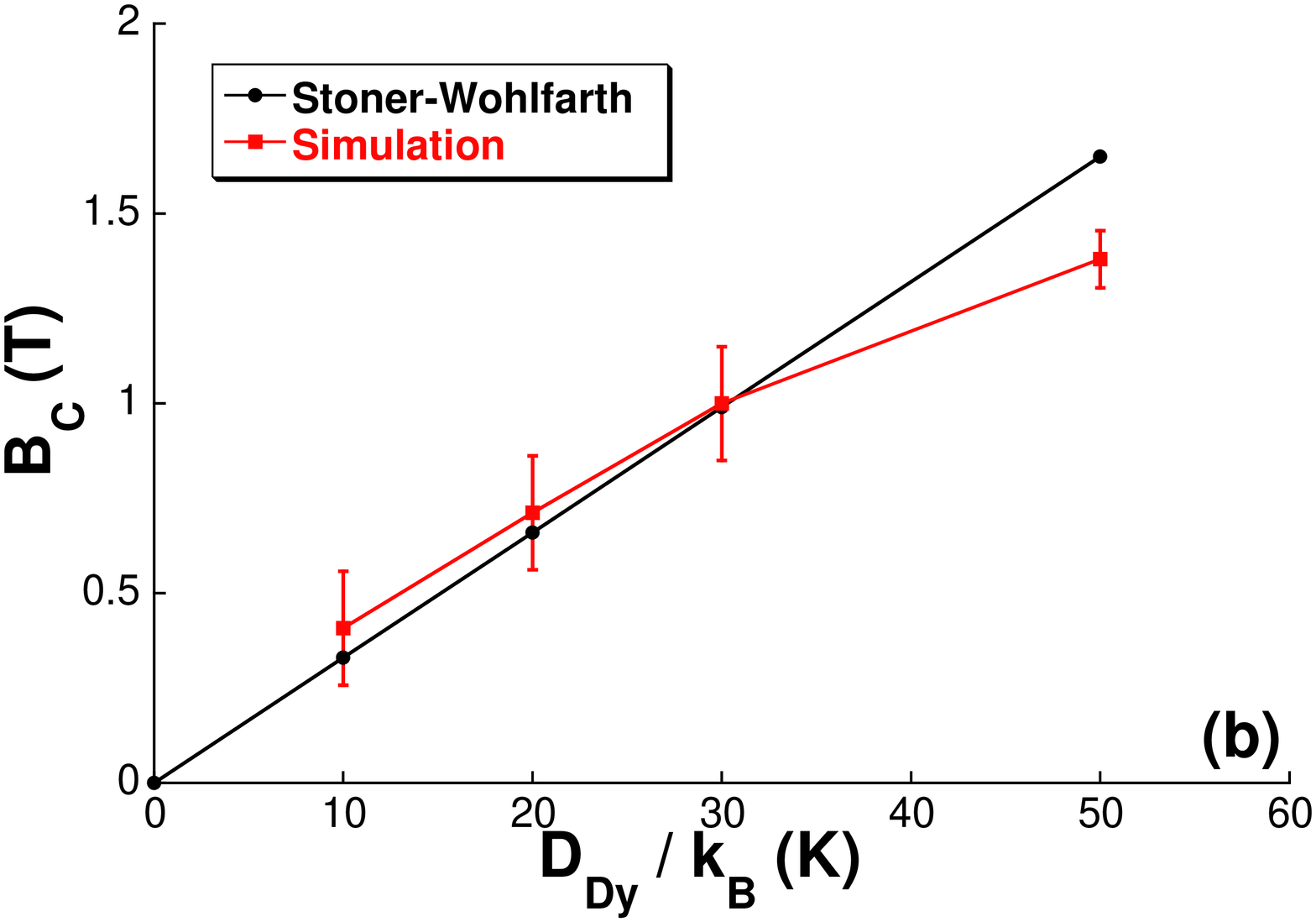}}
\caption{(a) Hysteresis loops in the perpendicular orientation for an Fe/Dy multilayer with an abrupt profile at $T=1$K and different values of the Dy anisotropy constant ; (b) Simulated and SW coercive field variation as a function of the Dy anisotropy constant value.}
\label{abrupt_amas}
\end{figure}

For the abrupt profile, the cluster concentration is fixed at 5\%; it cannot go further because the clusters are located at the interfaces. Fig.~\ref{abrupt_amas}.(a) shows the hysteresis loops for a magnetic field perpendicular to the film plane at $T=1$K and different values of $D_{\mathrm{Dy}}/k_{\mathrm{B}}$. We observe that the hysteresis loops are almost square indicating that the normal orientation is an easy axis. For all anisotropy constant values, the magnetic configuration is ferrimagnetic with the Dy moments being aligned with the external field. Unlike the case of uniaxial anisotropy examined in the previous section, we do not observe here IDW because the coercive field is always smaller than the spin-flop field ($\sim 2.1$T).

Fig.~\ref{abrupt_amas}.(b) shows the coercive field variation as a function of the Dy anisotropy constant value. Our results are in good agreement with the SW model up to $D_{\mathrm{Dy}}/k_{\mathrm{B}}=30$K which evidences a magnetization reversal by uniform rotation. Beyond this value, the reversal is no more purely uniform due to the clusters.

\subsection{APT profile}
\begin{figure}\centering
\rotatebox{-90}{\includegraphics[width=0.3\textwidth]{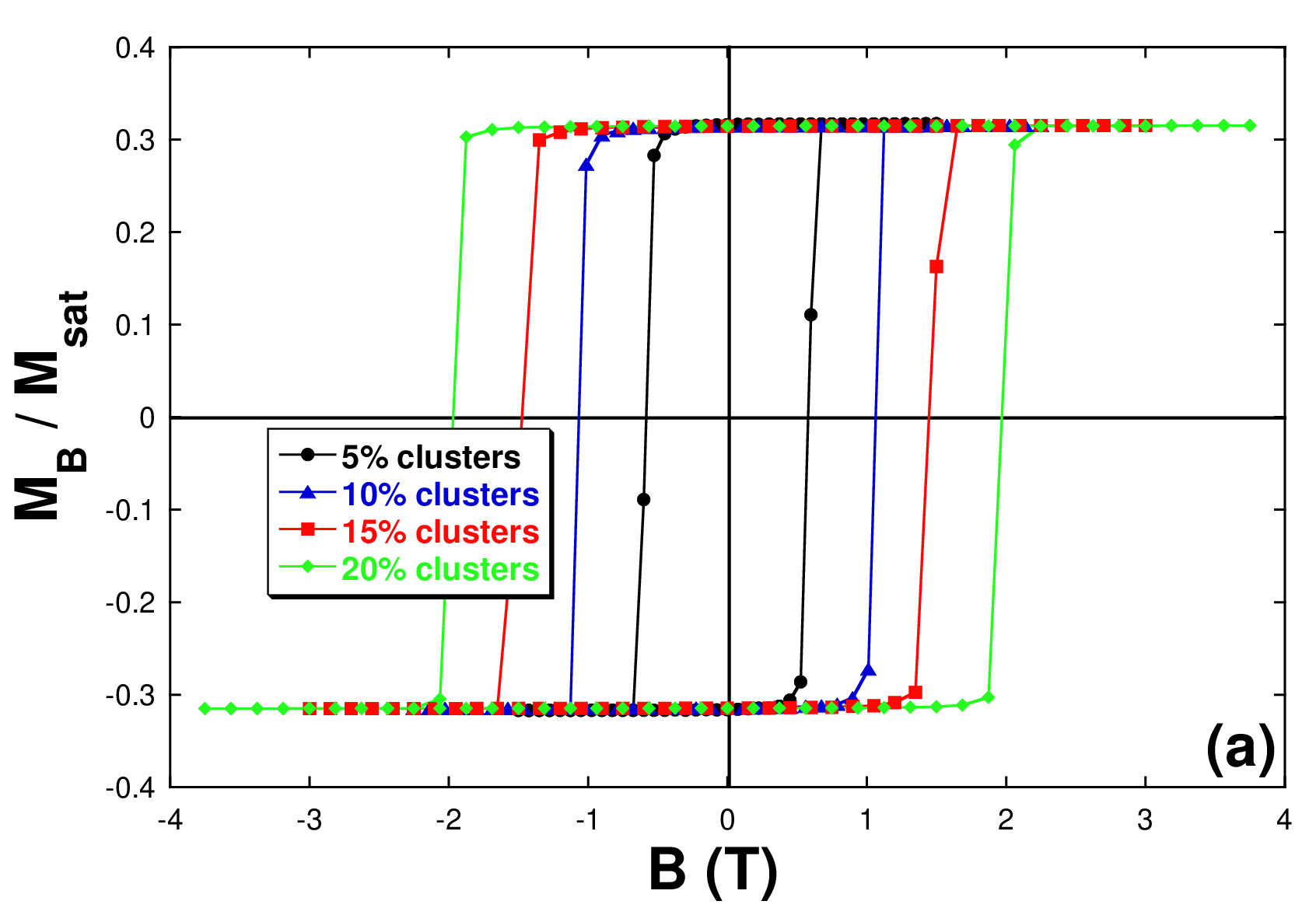}}
\rotatebox{-90}{\includegraphics[width=0.3\textwidth]{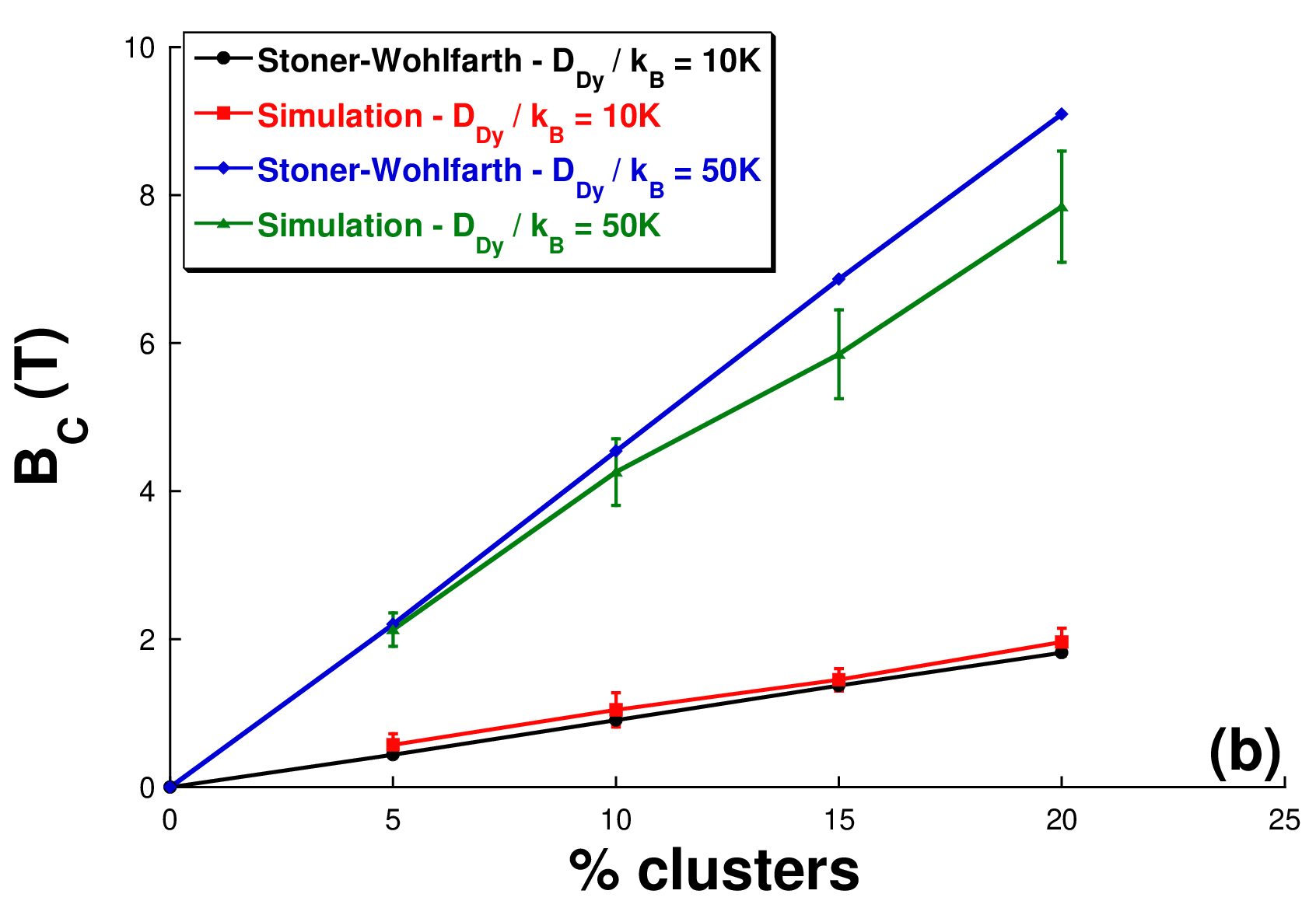}}
\caption{(a) Hysteresis loops in the perpendicular orientation for an Fe/Dy multilayer with an APT profile at $T=1$K and for $D_{\mathrm{Dy}}/k_{\mathrm{B}}=10$K and different cluster concentrations; (b) Simulated and SW coercive field variation as a function of the cluster concentration for $D_{\mathrm{Dy}}/k_{\mathrm{B}}=10$K and $D_{\mathrm{Dy}}/k_{\mathrm{B}}=50$K.}
\label{APT_amas}
\end{figure}

In the case of the APT profile, the clusters are not only localized at the interfaces. The hysteresis loops are represented on Fig.~\ref{APT_amas}.(a) at $T=1$K for $D_{\mathrm{Dy}}/k_{\mathrm{B}}=10$K and several cluster concentration values. As for the abrupt profile, our results indicate that the z-axis is an easy orientation and no IDW is observed for the same reason.

As can be seen in Fig.~\ref{APT_amas}.(b), for $D_{\mathrm{Dy}}/k_{\mathrm{B}}=10$K, the magnetization reversal process corresponds to an uniform rotation for all values of $c_{\mathrm{cluster}}$, whereas for $D_{\mathrm{Dy}}/k_{\mathrm{B}}=50$K the magnetization reversal is not purely uniform anymore starting from $c_{\mathrm{cluster}} \approx 15\%$.

The coercive field is proportional to the cluster concentration and to the Dy anisotropy constant as far as the product $c_{\mathrm{cluster}}D_{\mathrm{Dy}}/k_{\mathrm{B}}$ remains small enough ($\leq 5$K). We would like to note that, for a given anisotropy constant, the coercive field is slightly larger for an APT profile than for an abrupt profile (Tab.~\ref{table1}).

%

\section{\label{sec:amasrma} Cluster and random magnetic anisotropy}

In this section, we study a more realistic model of Fe/Dy multilayers to propose a qualitative explanation of the experimental hysteresis loops in terms of local magnetization reversal. This model takes into account an effective uniaxial anisotropy which is the result of the cluster anisotropy model and a random magnetic anisotropy (RMA) as proposed by Harris~\textit{et al}. \cite{HARR73} to investigate amorphous TM-RE compounds. Each Dy moment which is not inside a cluster, i.e. in the matrix, is assigned an uniaxial easy axis $\mathbf{n}_{i}$, these axes being randomly distributed. We have assumed that the anisotropy constant $D_{\mathrm{Dy}}$ is the same for all Dy atoms. The anisotropy energy term can then be written as:

\begin{equation}
	E_{a} = - D_{\mathrm{Dy}} \Big(  \sum_{i\in \text{Dy in matrix}} (\mathbf{S}_{i}.\mathbf{n}_{i})^{2} + \sum_{i \in \text{Dy in clusters}} (\mathbf{S}_{i}.\mathbf{z}_{i})^{2} \Big).
	\label{eq:Eaamas}
\end{equation}

\subsection{Abrupt profile}
\begin{figure}\centering
\rotatebox{-90}{\includegraphics[width=0.3\textwidth]{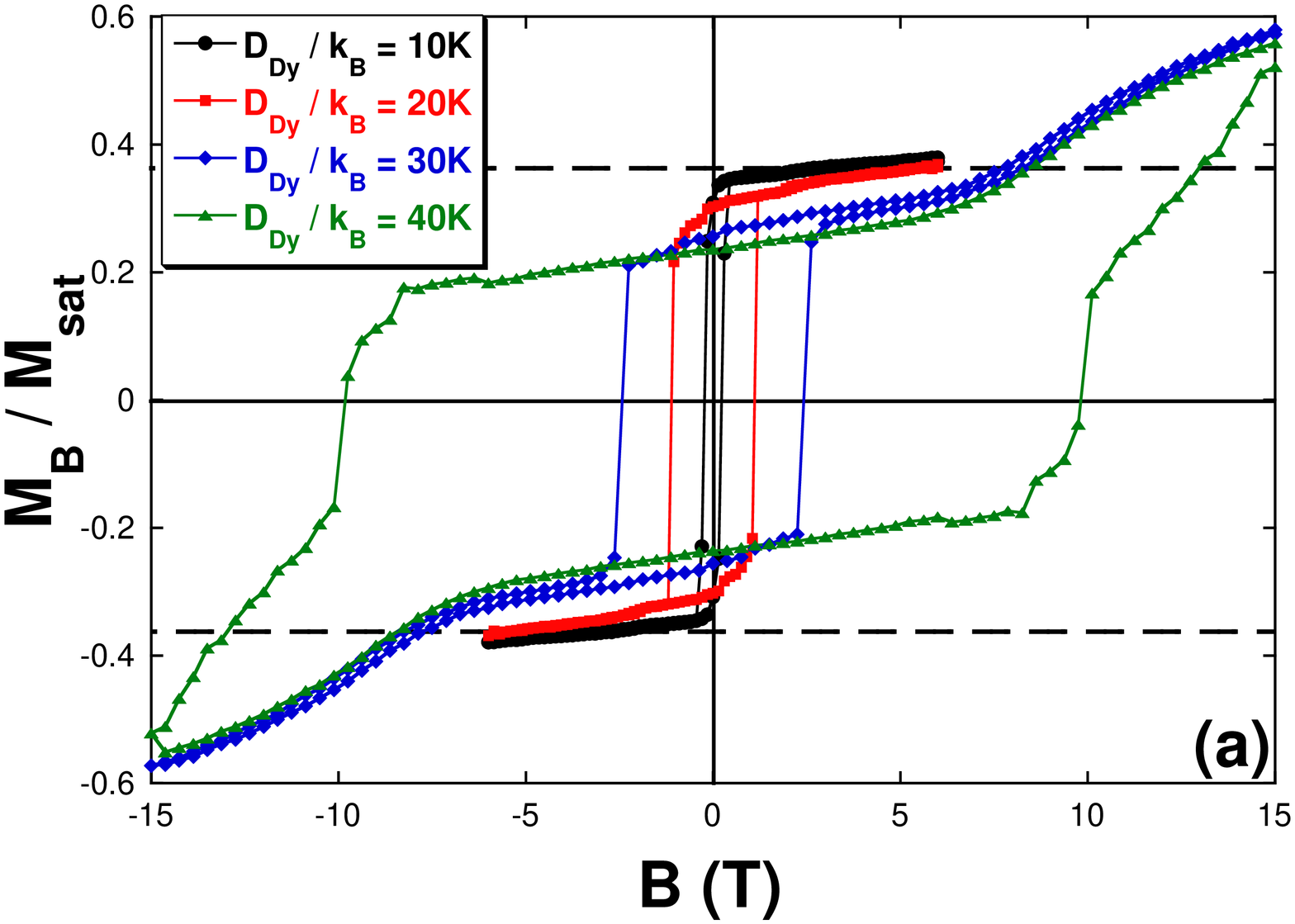}}
\rotatebox{-90}{\includegraphics[width=0.3\textwidth]{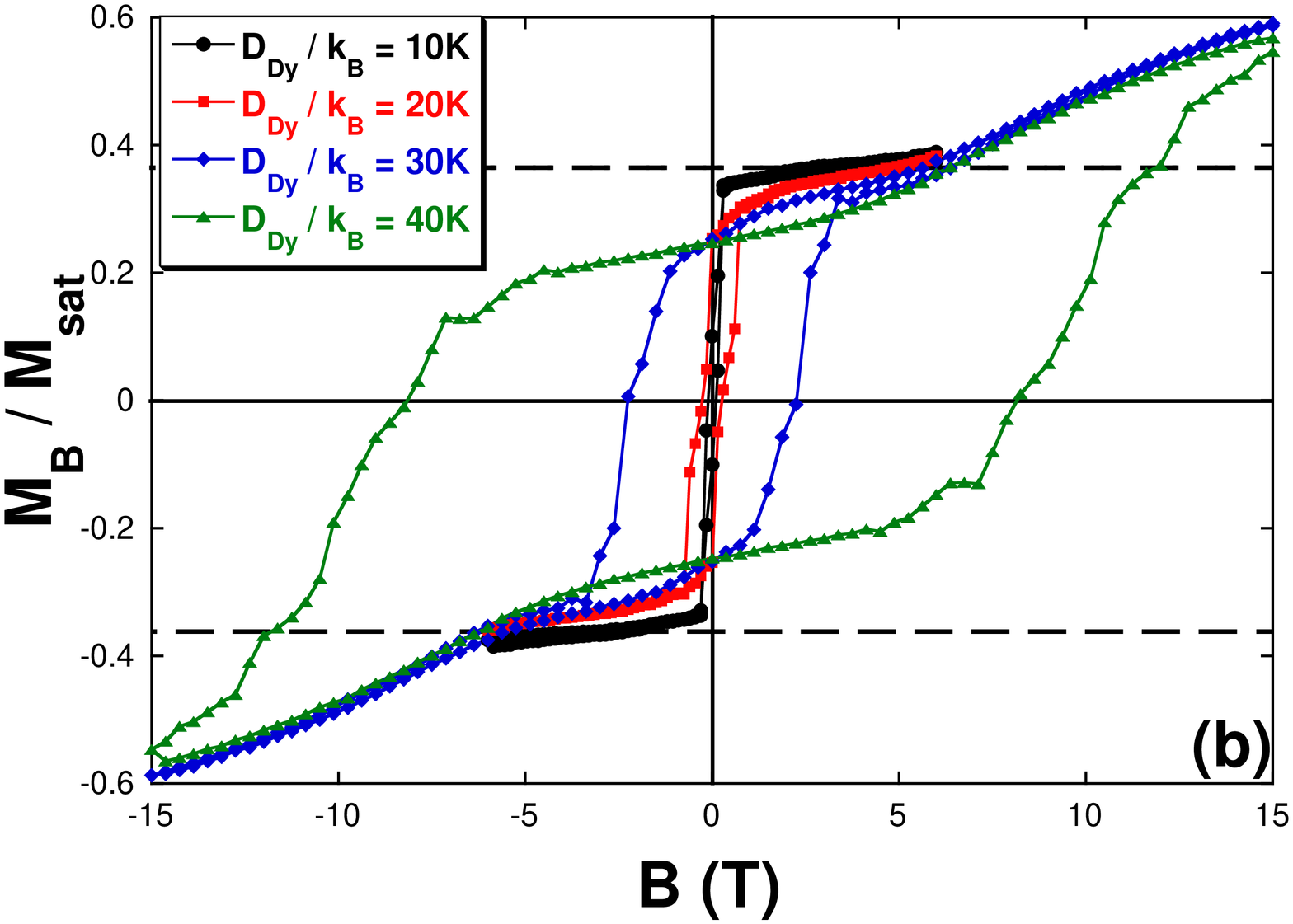}}
\caption{Hysteresis loops of an Fe/Dy multilayer with an abrupt profile, $c_{\mathrm{cluster}}=$5\% at $T=1$K for different Dy anisotropy constant values ((a) perpendicular; (b) parallel). The horizontal dashed lines correspond to $M_{\mathrm{ferrimagnetic}}/M_{\mathrm{sat}}$.}
\label{Abru_cycleamas}
\end{figure}

The hysteresis loops of the multilayers with an abrupt profile (the cluster concentration is equal to 5\%) for an in plane and normal applied field at $T=1$K are represented in Fig.~\ref{Abru_cycleamas}.

For $D_{\mathrm{Dy}} / k_{\mathrm{B}}=10$K and $20$K, the hysteresis loops are roughly those of a system with PMA: the loop for a perpendicular applied field is almost square and there is no remanence when the field is applied in the plane (see Table \ref{table2}). This behavior is due to the cluster anisotropy. The continous decrease of the magnetization before the reversal for a perpendicular field (Fig.~\ref{Abru_cycleamas}.(a)) is the result of an increase of sperimagnetism because of RMA (mainly in the Dy layers). The magnetization reversal mechanism is nearly uniform rotation in this case. For $D_{\mathrm{Dy}} / k_{\mathrm{B}}=30$K, no significant change is observed when the field is normal to the film except, of course, an increase of the coercive field. On the other hand, the loop is clearly open when the field is in the plane indicating a significant effect of the RMA.

%

\begin{figure}\centering
\rotatebox{-90}{\includegraphics[width=0.3\textwidth]{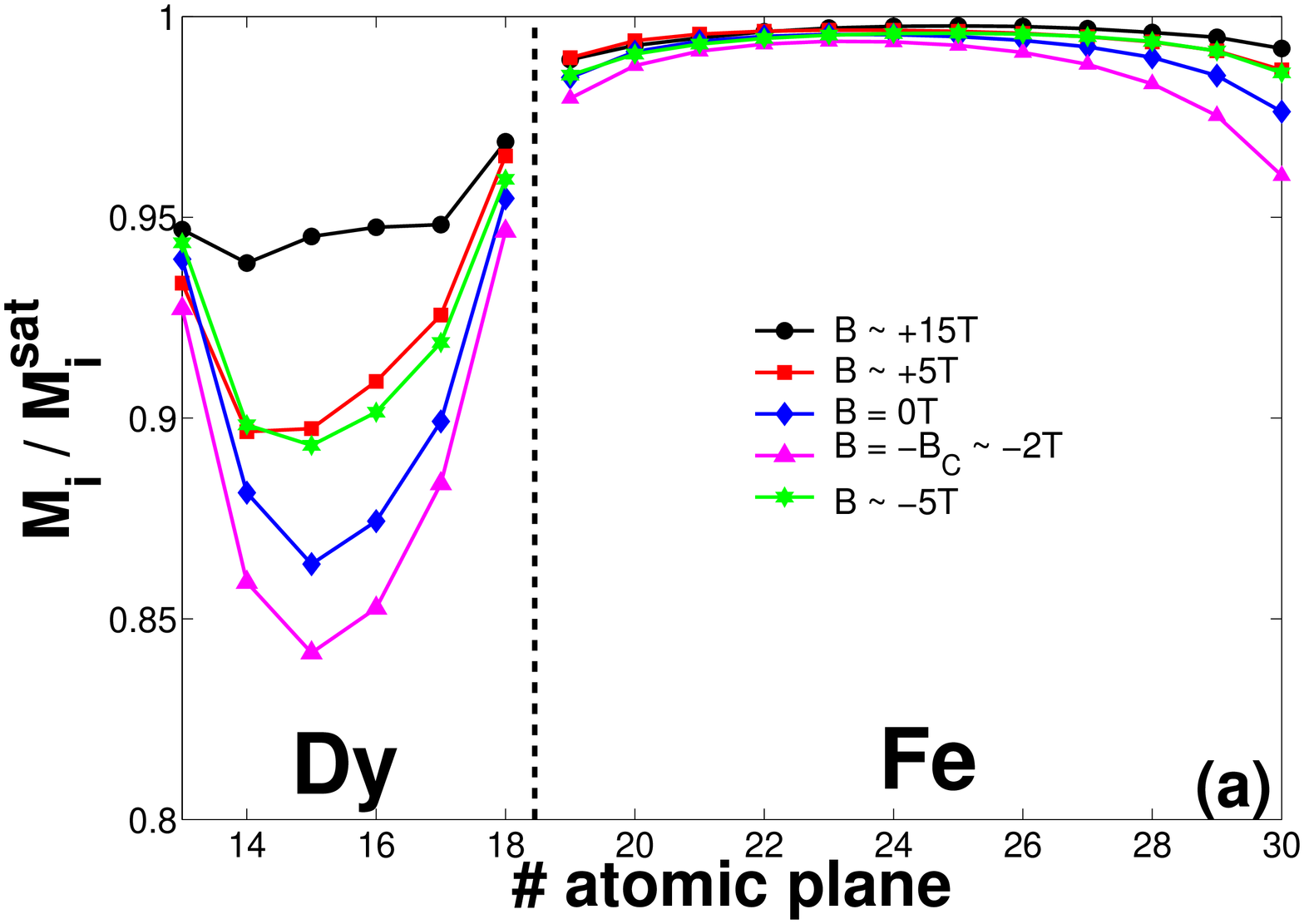}}
\rotatebox{-90}{\includegraphics[width=0.3\textwidth]{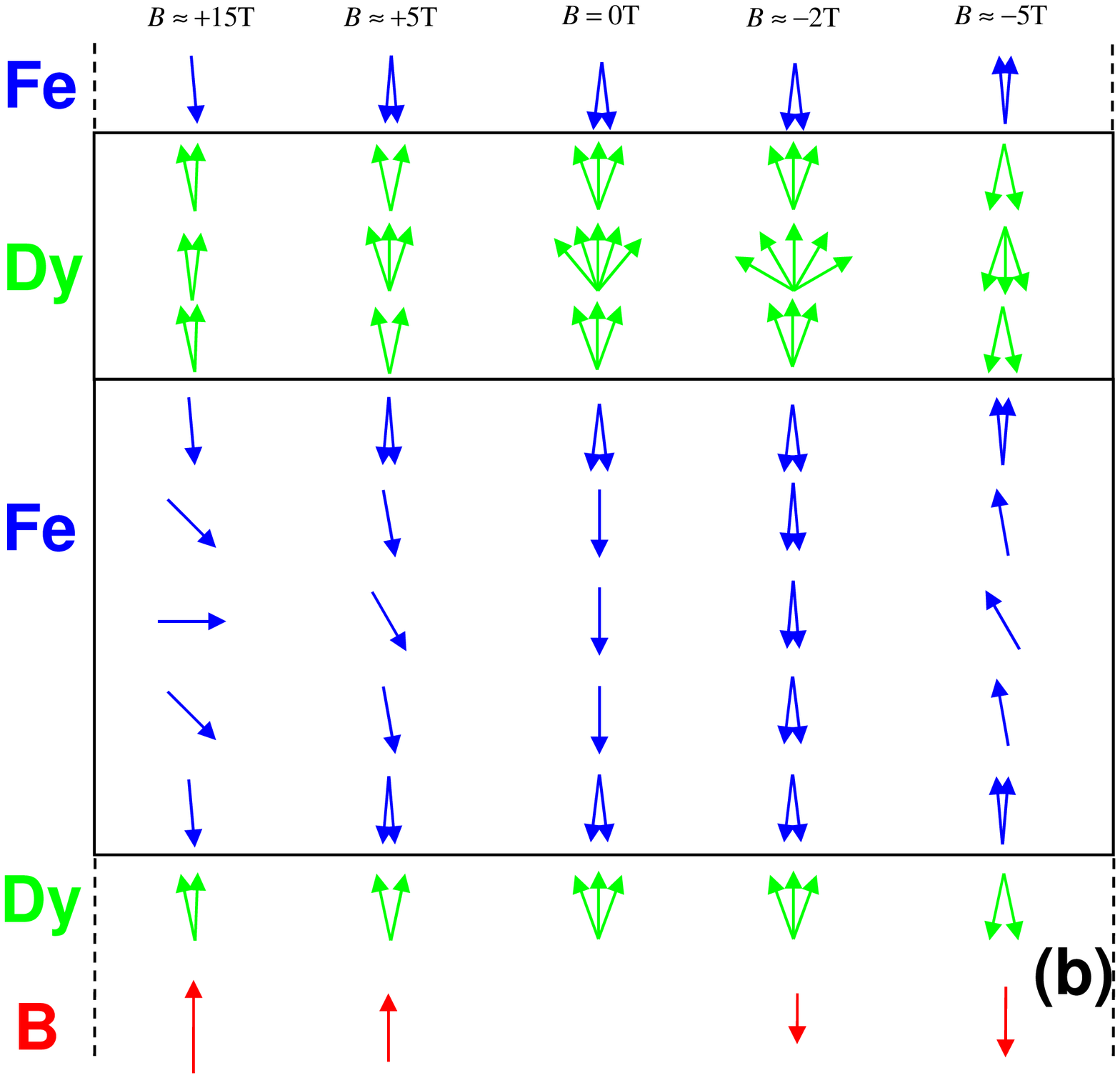}}
\caption{(a) Reduced magnetization of each atomic plane of a multilayer with an abrupt profile, $D_{\mathrm{Dy}}/k_{\mathrm{B}} = 30$K at $T=1$K for different magnetic field. (b) Corresponding schematic configurations. (The applied field is along the perpendicular orientation).}
\label{Abru_cycleamas2}
\end{figure}

The reduced magnetization profile and the corresponding magnetic configurations for different applied field (normal to the film) at $T=1$K are shown in Fig.~\ref{Abru_cycleamas2}.(a) and Fig.~\ref{Abru_cycleamas2}.(b), respectively. At $B=15$T, the ferrimagnetic order is broken in the core of the Fe layer which leads to a magnetization larger than $M_{\mathrm{ferrimagnetic}}$ as it can be seen in Fig.~\ref{Abru_cycleamas}.(a); this effect has not been observed experimentally since the applied magnetic field is not strong enough. Note that the Fe magnetic moments in each $xy$ plane are parallel whereas a small deviation from colinearity can be observed in the Dy planes ($M_{\mathrm{Dy}}/M_{\mathrm{sat}}^{\mathrm{Dy}} \sim 0.95$) (Fig.~\ref{Abru_cycleamas2}.(a)). As the field is reduced, down to 5T, two features are responsible for the decrease of the magnetization in Fig.~\ref{Abru_cycleamas}.(a): the Fe moments tend to be antiparallel to the Dy sublattice magnetization and the angular distribution of the Dy moments broadens (mainly in the core of the layer) (Fig.~\ref{Abru_cycleamas2}). The lowering of the remanent magnetization in the case of a perpendicular applied field compared to the cases $D_{\mathrm{Dy}} / k_{\mathrm{B}}=10$K and $20$K is clearly due to a more pronounced sperimagnetism. The dispersion of the moments is maximum just before the reversal ($B\sim -2$T in this case). After the reversal, the angular distribution of the moments sharpens as the field intensity increases and the average Fe moments in the core of the layer deviate from the $z$ axis. For $D_{\mathrm{Dy}} / k_{\mathrm{B}}=40$K, the hysteresis loops for a field in the plane or normal to it are quite similar showing that the magnetization reversal is, this time, essentially governed by the RMA and exchange interactions. This is consistent with the coercive field values for the two orientations reported in table.~\ref{table2}. Indeed the coercive field for an in-plane or perpendicular applied field are very close for $D_{\mathrm{Dy}}/k_{\mathrm{B}} \geq 30$K within the uncertainties which are significant in this case owing to strong competition between RMA and exchange interactions. So, our investigation evidences a crossover from an uniaxial anisotropy (PMA) type behavior at low $D_{\mathrm{Dy}}$ values to a RMA type behavior at high $D_{\mathrm{Dy}}$ values. Of course, in this latter case, the magnetization reversal mechanism is clearly non uniform. In order to explain quantitatively this crossover, we have estimated the exchange and anisotropy contributions to the local field on a Dy site in the matrix. These two contributions are respectively given by:

\begin{equation}
	\mu_{0} \mathbf{H}_{\mathrm{Dy}}^{\mathrm{e}} = \frac{1}{2} J_{\mathrm{Dy-Dy}} \frac{(g_{\mathrm{Dy}}-1)^{2}}{(g_{\mathrm{Dy}} \mu_{\mathrm{B}})^{2}} \sum_{j} \mathbf{m}_{\mathrm{Dy}}^{j}	,
\label{eq:chechv}
\end{equation}
\begin{equation}
	\mu_{0} \mathbf{H}_{\mathrm{Dy}}^{\mathrm{a}} = \frac{2 D_{\mathrm{Dy}} (g_{\mathrm{Dy}}-1)^{2}}{(g_{\mathrm{Dy}} \mu_{\mathrm{B}})^{2}} (\mathbf{m}^{i}_{\mathrm{Dy}}.\mathbf{n}_{i})\mathbf{n}_{i}.	
\label{eq:chanv}
\end{equation}

Assuming that the Dy moments are almost parallel to the $z$ axis, we have $\mu_{0}H_{e}=(5ZJ_{\mathrm{Dy-Dy}})/(16 \mu_{\mathrm{B}}) \sim 36.5$T ($Z=12$) and $\mu_{0}H_{a}=(5D_{\mathrm{Dy}})/(8 \mu_{\mathrm{B}}$) (using $<\mathbf{m}^{i}_{\mathrm{Dy}}.\mathbf{n}_{i}>=m_{\mathrm{Dy}}/2$ where $<>$ means average over the anisotropy axes). The average value of the anisotropy contribution varies from 9.3T to 37.2T. Actually, these two contributions are of the same order of magnitude when $D_{\mathrm{Dy}} / k_{\mathrm{B}}\sim30$K which is consistent with the observed change in the magnetization reversal mechanism.

We would like to emphasize that RMA increases significantly the coercive field values for large enough anisotropy constant $D_{\mathrm{Dy}}$ in comparison with the cluster anisotropy model (Sec.~\ref{sec:amas}). Indeed, for small values of $D_{\mathrm{Dy}}$, the magnetization reversal is roughly uniform and the RMA effect is very small. On the other hand, increasing $D_{\mathrm{Dy}}$ leads to a non-uniform reversal process. Then, each magnetic moment is individually sensitive to the local anisotropy field which is due to RMA on the Dy sites of the matrix.

 Finally, let us mention that although the fraction of Dy atoms in the clusters relatively to the total number of Dy atoms is constant (3.4\%), it is quite surprising that the multilayer behaves either as an uniaxial anisotropy system ($D_{\mathrm{Dy}}$ small) or a RMA system ($D_{\mathrm{Dy}}$ large). In the limit of very large anisotropy constant values, the small crystallized clusters would have no effect.

\subsection{APT profile}
\begin{figure}\centering
\rotatebox{-90}{\includegraphics[width=0.3\textwidth]{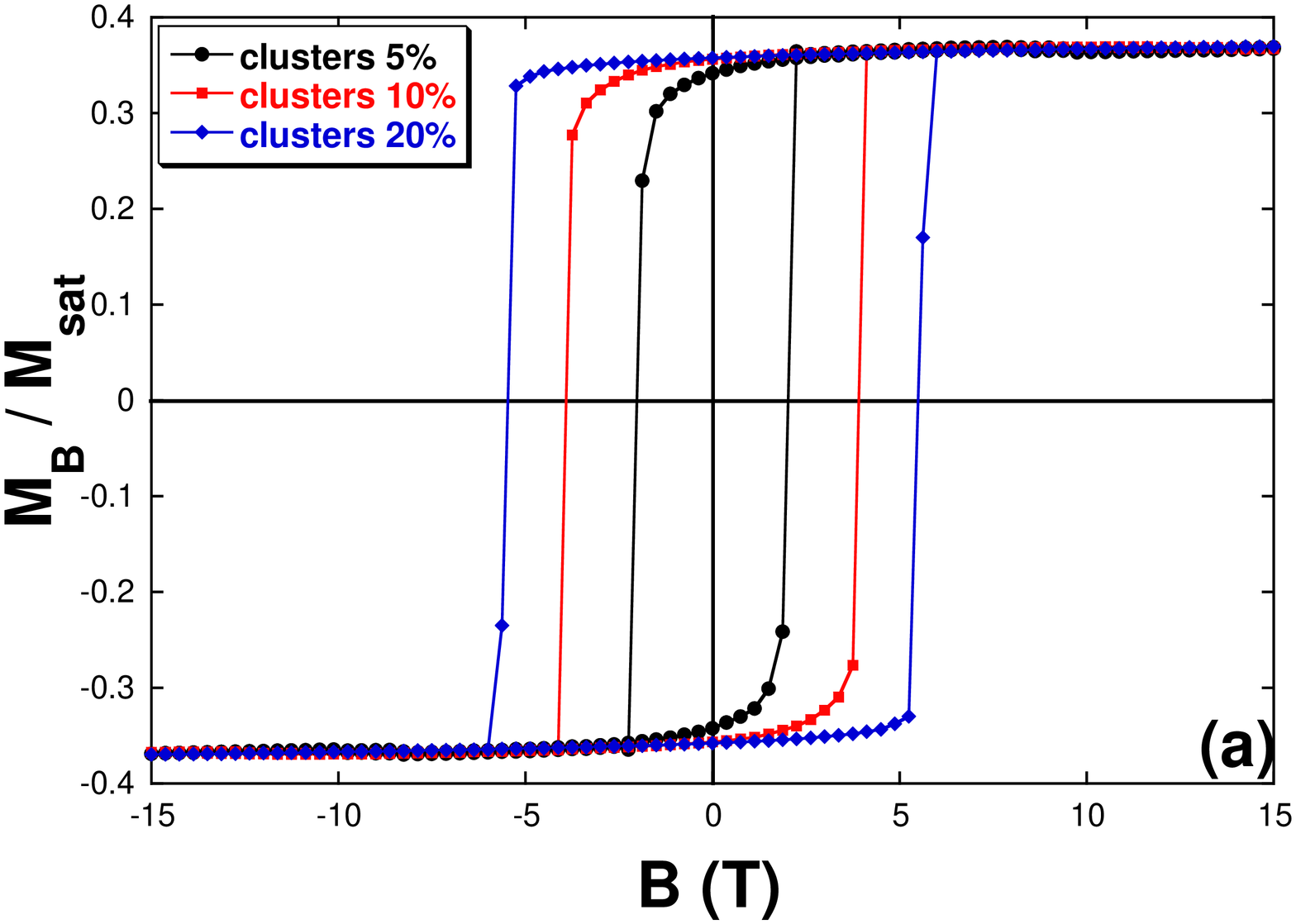}}
\rotatebox{-90}{\includegraphics[width=0.3\textwidth]{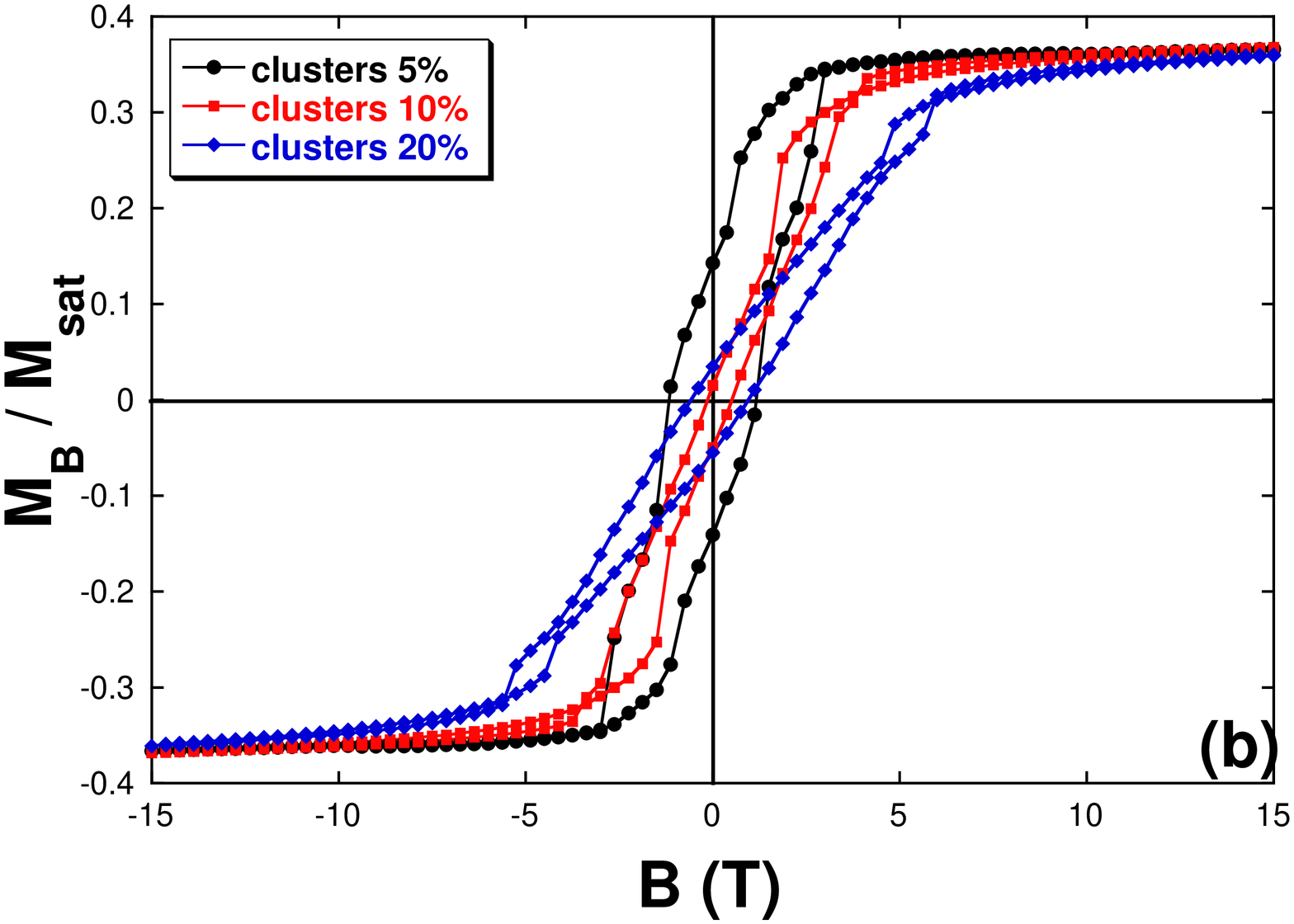}}
\caption{Hysteresis loops of an Fe/Dy multilayer with an APT profile at $T=1$K and $D_{\mathrm{Dy}}/k_{\mathrm{B}} = 30$K for different values of the cluster concentration ((a) perpendicular; (b) parallel).}
\label{Sonde_cycleamas}
\end{figure}

The case of the APT profile is very interesting since it allows to investigate the influence of the cluster concentration. In Fig.~\ref{Sonde_cycleamas}, we have shown the hysteresis loops for different cluster concentrations at $T=1$K and for $D_{\mathrm{Dy}}/k_{\mathrm{B}} = 30$K. The main effect of the cluster concentration is to enhance the coercive field. The hysteresis loops in Fig.~\ref{Sonde_cycleamas}.(a) indicate an easy axis along the normal to the film plane. The small decrease of the magnetization before the reversal (for $c_{\mathrm{cluster}}= 5\%$ and 10\%) is attributed to the increase of the dispersion of the moments owing to the RMA. This effect tends to disappear as the cluster concentration is increased since this latter favors uniform rotation of the spins.

When the field is applied in the film plane (Fig.~\ref{Sonde_cycleamas}.(b)), we observe that the uncertainties are quite large ($\Delta B_{\mathrm{C}} \sim 0.5$T) since the coercive field does not vary monotonously with $c_{\mathrm{cluster}}$ as it should be. Consequently, no significant effect of the cluster concentration is noted. However, our results indicate that the coercive field values are non zero due to the RMA but are much smaller than in the perpendicular orientation. As for the abrupt profile, we have observed that the magnetic moment dispersion is maximum just before the magnetization reversal (not shown here).

\subsection{Discussion}
Our results evidence that the APT profile favors perpendicular anisotropy in comparison to the abrupt profile (see Figs.~\ref{Abru_cycleamas} and \ref{Sonde_cycleamas}). Concerning the abrupt profile, we observe a PMA behavior only when the Dy anisotropy constant value is lower than $30$K. Experimentally, the hysteresis loops measured on the multilayers elaborated at low temperature, with a concentration profile close to an abrupt profile do not display any PMA unlike the sample built up at $T=570$K. It means that our simulation model can be qualitatively compared to the experimental case when the anisotropy constant value $D_{\mathrm{Dy}}/k_{\mathrm{B}}$ is larger than $30$K. For $D_{\mathrm{Dy}}/k_{\mathrm{B}} = 30$K and $c_{\mathrm{cluster}}= 5\%$, the multilayer with an APT profile exhibits roughly uniaxial magnetic system hysteresis loops ($B_{\mathrm{C}}^{\perp} \sim 2.00 \pm 0.15$T) contrary to the abrupt profile multilayer. One should note that the coercive field is slightly larger for the abrupt profile ($B_{\mathrm{C}}^{\perp} \sim 2.45 \pm 0.15$T) unlike the case without RMA (see Sec.~\ref{sec:amas}). This can be explained as following: as previously mentioned \cite{TALB08} the sperimagnetism due to RMA is more pronounced for the abrupt profile, so the increase of $B_{\mathrm{C}}$ when adding RMA is larger than for the APT profile.

\section{\label{sec:disc} Influence of the temperature}
\begin{figure*}\centering
\rotatebox{-90}{\includegraphics[width=0.3\textwidth]{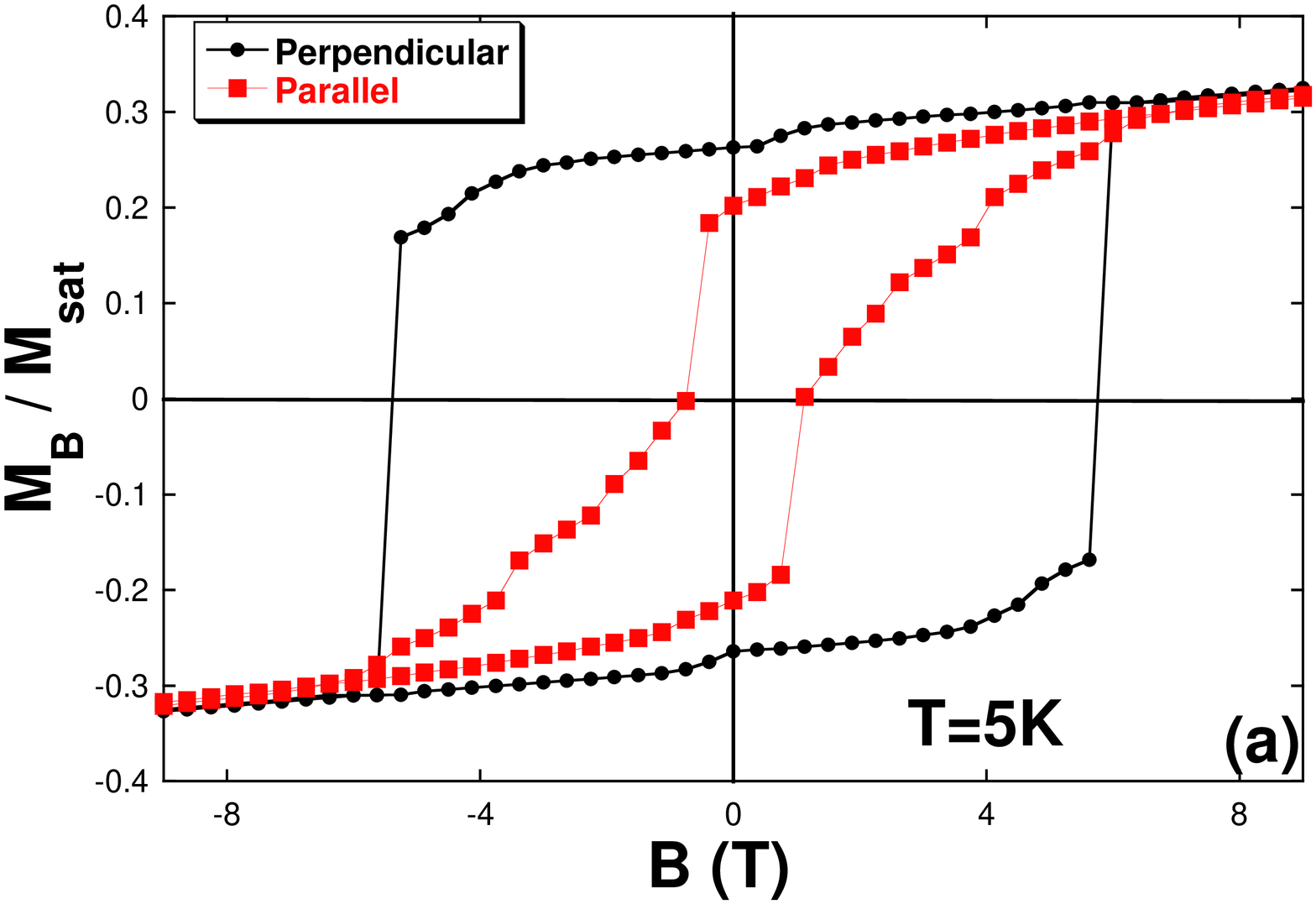}}
\rotatebox{-90}{\includegraphics[width=0.3\textwidth]{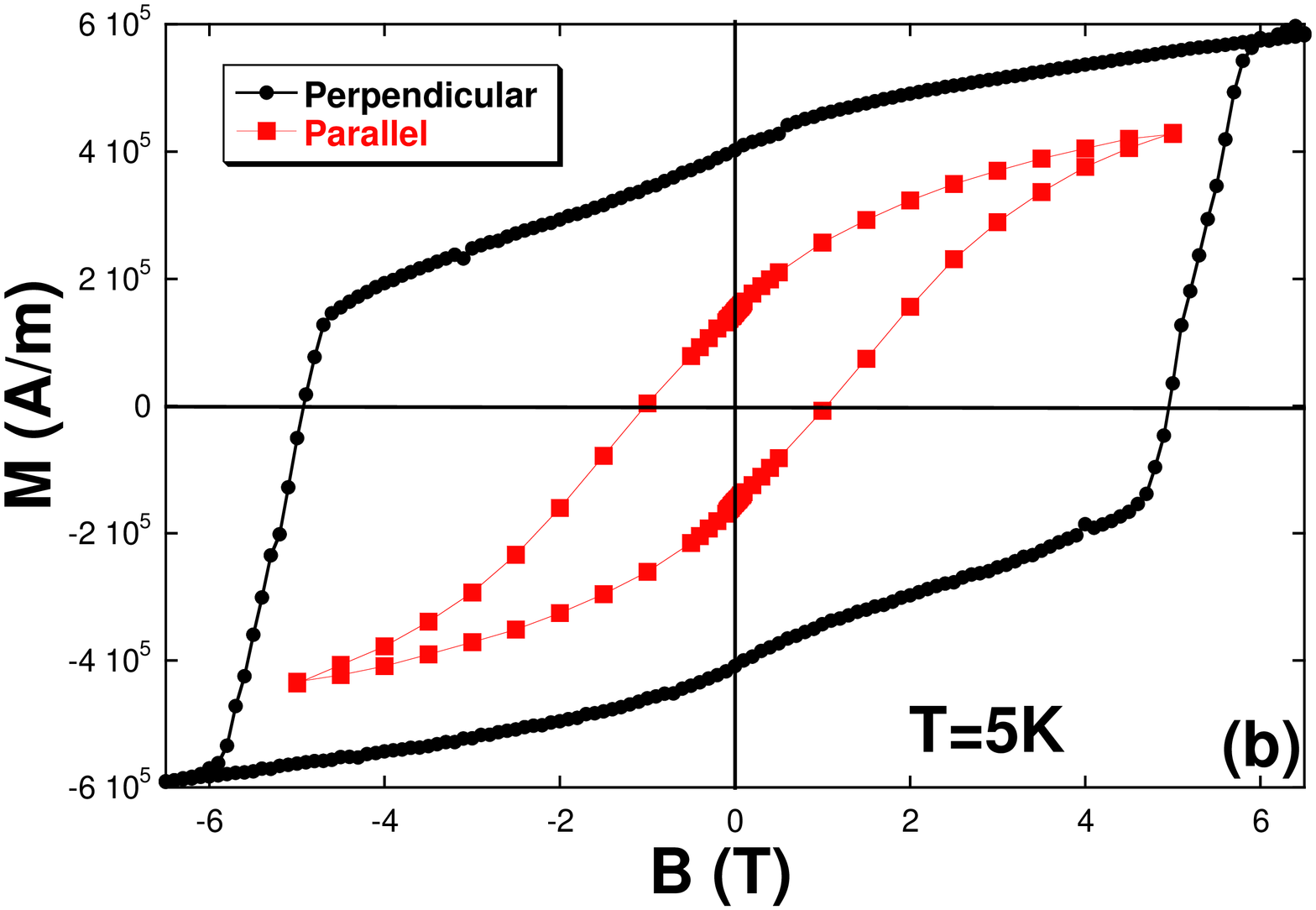}}
\rotatebox{-90}{\includegraphics[width=0.3\textwidth]{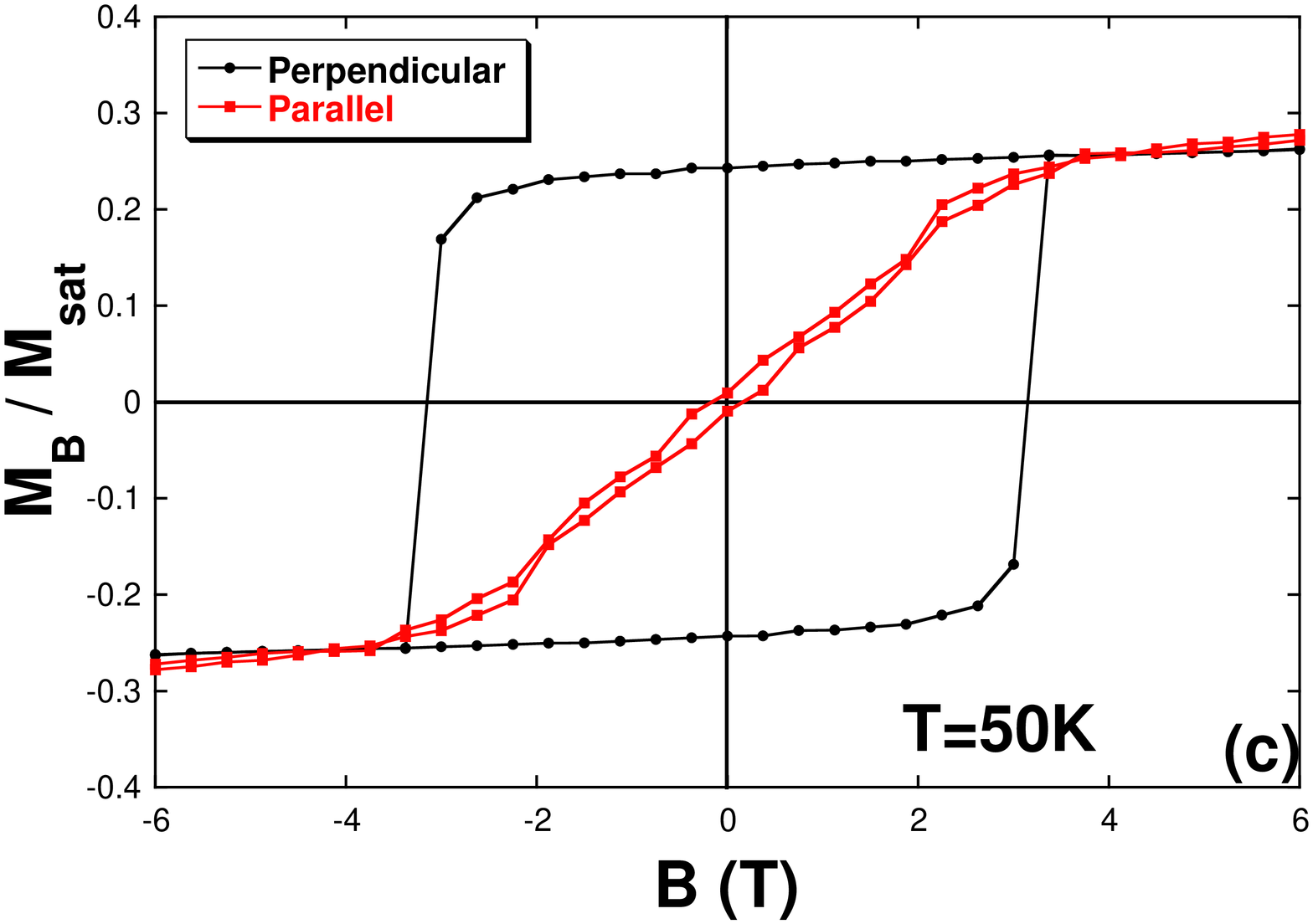}}
\rotatebox{-90}{\includegraphics[width=0.3\textwidth]{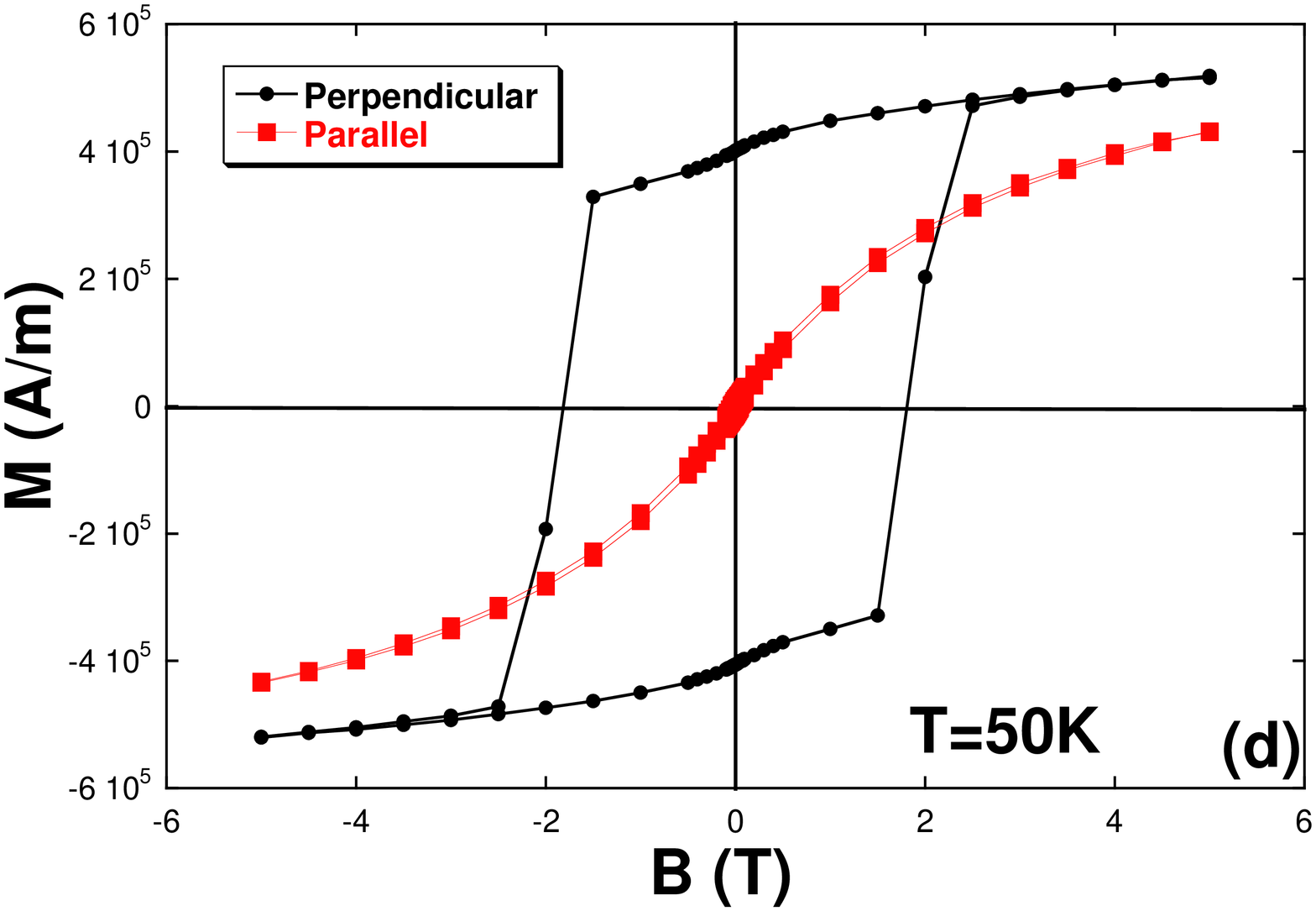}}
\rotatebox{-90}{\includegraphics[width=0.3\textwidth]{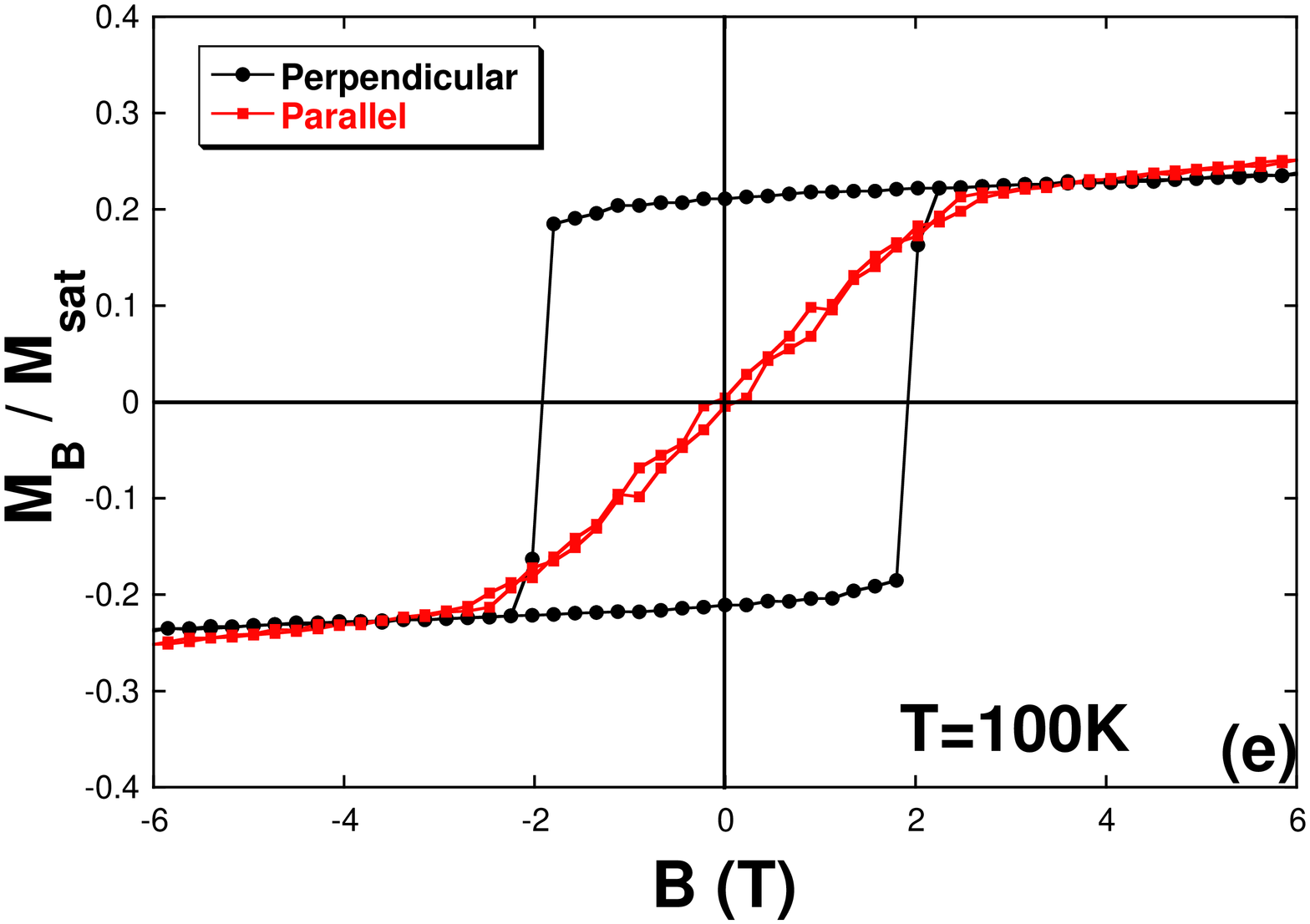}}
\rotatebox{-90}{\includegraphics[width=0.3\textwidth]{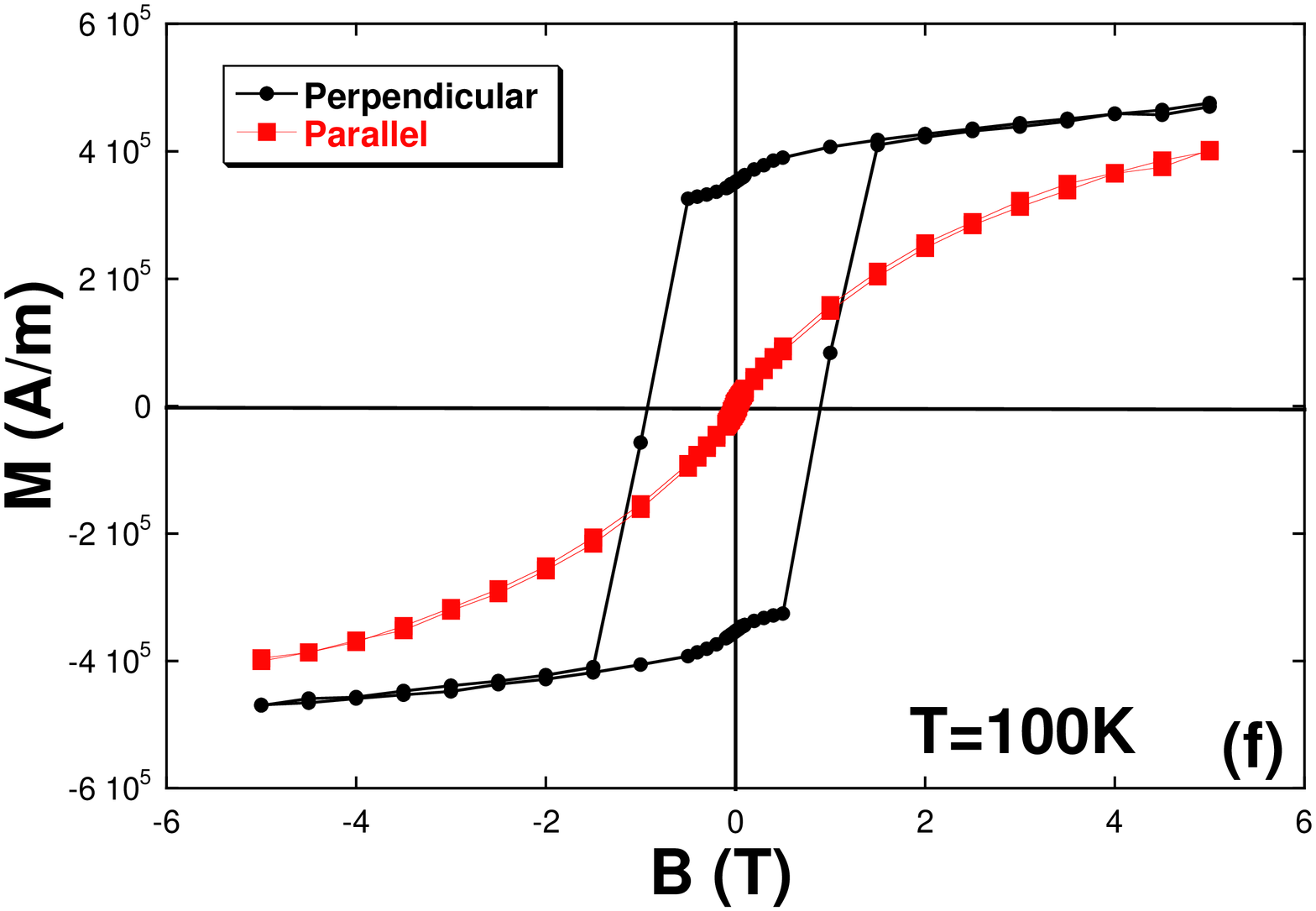}}
\caption{Hysteresis loops of Fe/Dy multilayers at $T=5,50,100$K. (a, c, e) simulated loops with an APT profile, $D_{\mathrm{Dy}}/k_{\mathrm{B}} = 40$K and $c_{\mathrm{cluster}}=$10\%; (b, d, f) experimental loops for multilayer elaborated at $570$K.}
\label{Sonde_cycleamasT}
\end{figure*}

The temperature effect has been investigated in the case of the APT profile and compared to the experimental results of an (Fe 3nm/Dy 2nm) multilayer elaborated at 570K. The simulated and experimental hysteresis loops at $T=5$, 50 and 100K are shown in Fig.~\ref{Sonde_cycleamasT}. The numerical results have been obtained with $D_{\mathrm{Dy}} / k_{\mathrm{B}}= 40$K and $c_{\mathrm{cluster}}=$10\% for which the best agreement with experimental data at low temperature has been observed. At $T=5$K, the multilayer exhibits PMA unlike the multilayer with an abrupt profile (previous section). Note that the loop with an in-plane field at $T=5$K is open contrary to the case $D_{\mathrm{Dy}} / k_{\mathrm{B}}= 30$K (previous section).

The temperature dependence of the simulated loops is in good qualitative agreement with the experimental ones. For each temperature, the hysteresis loops evidence an easy axis normal to the film. The coercive field decreases as the temperature increases because of the lowering of the effective anisotropy constant in relation with thermal fluctuations of the atomic moments. The hysteresis loops with an applied field in the film plane are open at low temperature owing to RMA. Above $50$K, no hysteretic phenomenon is observed indicating that RMA has no effect anymore in this temperature range. The hysteresis loops of our model above $50$K are those of a uniaxial anisotropy type system with an easy orientation along the $z$ axis. It can be seen that unlike the simulated loops, the experimental ones do not saturate mainly for an applied field in the film plane. This indicates that the fan structure of the Dy moments is more rigid in real systems than in our model.


\section{\label{sec:concl} Conclusion}

In this work, we have investigated a model of amorphous Fe/Dy multilayers with RMA and small crystallized clusters defining on average a preferential axis normal to the film plane. Our results using a model built up from experimental concentration profiles are consistent with measured hysteresis loops, i.e. APT profile favors PMA in comparison to the abrupt profile. It has been shown that the crystallized clusters embedded in the matrix with a very small concentration could explain the observed PMA. Moreover, the cluster anisotropy effect is more pronounced with the APT profile. With the abrupt profile, RMA plays a more significant role and may even hide the cluster anisotropy effect when the anisotropy constant is enlarged. The temperature influence on a multilayer with an APT profile is to remove RMA effect. In a near future, we plan to investigate magnetostrictive and magnetoresistive multilayers.

\begin{acknowledgments}
The simulations were performed at the Centre de Ressources Informatiques de Haute Normandie (CRIHAN) under the project No. 2004002. Moreover, the authors are indebted to Dr. Alexandre Tamion and Dr. Catherine Bordel for valuable discussions
on the experimental results.
\end{acknowledgments}

\clearpage
\begin{table}[h]
\caption{Coercive field values at $T=1$K of the multilayer with the abrupt and the APT profiles with $c_{\mathrm{cluster}}=5\% $ for $D_{\mathrm{Dy}} / k_{\mathrm{B}} = 10$K and $50$K.}
\label{table1}
\begin{ruledtabular}
\begin{tabular}{ccc}
 $D_{\mathrm{Dy}}/k_{\mathrm{B}} $ (K) & $B_{\mathrm{C}}$ (T) Abrupt & $B_{\mathrm{C}}$ (T) APT\\
\hline
$10$ & $0.41 \pm 0.15$ & $0.57 \pm 0.15$ \\

$50$ & $1.38 \pm 0.15$ & $2.13 \pm 0.15$ \\
\end{tabular}
\end{ruledtabular}
\end{table}

\clearpage

\begin{table}[h]
\caption{Coercive field values at $T=1$K of the multilayer with the abrupt profile with $c_{\mathrm{cluster}}=5\% $ for various values of $D_{\mathrm{Dy}} / k_{\mathrm{B}}$ in perpendicular and parallel field orientations.}
\label{table2}
\begin{ruledtabular}
\begin{tabular}{ccccc}
$\frac{D_{\mathrm{Dy}}}{k_{\mathrm{B}}}$ (K) & $10$ & $20$ & $30$ & $40$\\
 \hline
$B_{\mathrm{C}}^{\perp}$ (T) & $0.33 \pm 0.15$ & $1.10 \pm 0.15$  & $2.45 \pm 0.30$  & $9.80 \pm 0.50$ \\

$B_{\mathrm{C}}^{//}$ (T) & $0.10 \pm 0.15$ & $0.20 \pm 0.15$ & $2.20 \pm 0.30$  & $8.40 \pm 0.50$ \\
\end{tabular}
\end{ruledtabular}
\end{table}
\clearpage

\end{document}